# Nonperturbative renormalization group approach to turbulence

A. Esser* and S. Grossmann†

*Fachbereich Physik der Philipps–Universität, Renthof 6, D-35032, Marburg, Germany*

(October 5, 2018)

## Abstract

We suggest a new, renormalization group (RG) based, nonperturbative method for treating the intermittency problem of fully developed turbulence which also includes the effects of a finite boundary of the turbulent flow. The key idea is not to try to construct an elimination procedure based on some assumed statistical distribution, but to make an ansatz for possible RG transformations and to pose constraints upon those, which guarantee the invariance of the nonlinear term in the Navier–Stokes equation, the invariance of the energy dissipation, and other basic properties of the velocity field. The role of length scales is taken to be inverse to that in the theory of critical phenomena; thus possible intermittency corrections are connected with the outer length scale. Depending on the specific type of flow, we find different sets of admissible transformations with distinct scaling behavior: for the often considered infinite, isotropic, and homogeneous system K41 scaling is enforced, but for the more realistic plane Couette geometry no restrictions on intermittency

---

*present address: Nokia Telecommunications, 00045 Nokia Group, Finland, e-mail: Alexander.Esser@ntc.nokia.com

†e-mail: grossmann_s@physik.uni-marburg.de

exponents were obtained so far.

# I. INTRODUCTION

A fundamental, yet unsolved problem of fully developed turbulence is the scaling behaviour of the structure functions in the inertial range for Reynolds numbers approaching infinity.

Although many if not most experiments seem to favour the existence of intermittency corrections [1–5], some other experiments seem to confirm the classical "K41" exponents [6,7]. Thus, the possibility remains that measured intermittency corrections are due to still–to–small Reynolds numbers, invalidity of Taylor's hypothesis, lack of statistics or other experimental deficiencies.

On the theoretical side, there hasn't been much progress in this regard either since K41 theory [8–12] and mean–field theories such as Ref. [13] and Kraichnan's LHDIA [14]. Numerical simulations [15–17] still do not achieve high enough Reynolds numbers, and approximations allowing higher Reynolds numbers such as REWA [18–20] reduce the degrees of freedom without control of the consequences. Phenomenological models [21–33] are not only limited to physically "explaining" intermittency corrections without being able to connect intermittency with the Navier–Stokes dynamics, but cannot even demonstrate its existence for Navier–Stokes turbulence since statistical models generically show intermittency anyhow [34].

The basic motivation to analyze turbulence with RG methods comes from a couple of similarities with critical phenomena: the importance of extremely many spatial degrees of freedom, strong fluctuations, and nonlinear interaction, to name a few. These ingredients are essential for spatially extended self-similar structures and scaling laws with nontrivial exponents. We now briefly discuss some of the many efforts to derive inertial range scaling exponents with the aid of RG and/or field theory; current overviews can be found in Ref. [35–38].

The first, by now "classical" approach of Ref. [39–41] formally adopted the formalism of Wilson style respectively field-theoretic RG theory for critical phenomena to hydrodynamics.

Microscopic degrees of freedom are eliminated by averaging over an artificial, stochastic, and microscopic volume forcing. The intermittency exponents depend on the forcing's spectral strength, and these authors explicitly state that such a theory is not adequate for fully developed turbulence. Nevertheless there are several extensions of these theories such as Ref. [42,43] where the incorporation of traditional closure methods enables the calculation of dimensionless numbers like Kolmogorov's constant. Other work employs a driving force restricted to small wave numbers without being able to calculate an effective, RG invariant forcing [44]. Yet another idea is to examine freely decaying turbulence with random initial conditions [45]. Last but not least, formal analogies to other RG theories such as renormalized quantum field theory are possible [37].

Quite early in the game, another mapping to critical phenomena was suggested which seems to match the common phenomenological point of view of turbulence much better [46–50]. It is "inverse" in the sense that macroscopic degrees of freedom should be eliminated, and that, consequently, intermittency corrections are related to the outer length scale. Another, within this picture obvious, but actually rarely suggested idea is to take the energy dissipation rate to be the RG invariant quantity instead of the free energy which does not seem to have any special role in turbulence [48,51]. We are aware of two actual attempts for an inverse RG theory, one with Wilson-style elimination of degrees of freedom [48] and one field-theoretic one [52]. Although being inverse, both adopt major basic properties of thermodynamic systems which do not seem to be adequate for turbulence.

All methods mentioned so far use perturbation theory with the strength of the nonlinear interaction term of the Navier–Stokes equation being treated as a formal expansion parameter which is problematic without a suitable resummation procedure. Furthermore there are unphysical infrared divergencies; a problem which can be solved by using a formal perturbation approach with velocity differences instead of Eulerian velocities [53]. This perturbation theory is still being worked on [54–56], but so far not with RG methods.

Without judging all these efforts, it is fair to say that the calculation of inertial range scaling exponents from the Navier–Stokes equation has not been achieved yet.

There is a very important aspect of turbulence which is mostly neglected in turbulent theories: the boundaries of the system. Numerical simulations usually employ periodic boundary conditions and analytical theories try to take care of the finite system size with a formal "outer length" parameter. Energy input, which is required to balance energy dissipation, is realized with a volume forcing which might be unphysical.

In this work now, we introduce a new, analytical and inverse RG method which is based on the Navier–Stokes equation and tries to take into account the boundaries. The paper is organized as follows: in Sect. II we derive the equation of motion for certain velocity differences, which will be our order parameter. In Sect. III we explain the physical ideas of the new RG approach before actually calculating RG transformations in Sect. V–VII. We end with a summary and discussion in Sect. VIII.

## II. EQUATIONS OF MOTION

Our starting point is the Navier–Stokes equation

$$\partial_t \mathbf{u} + \mathbf{u} \cdot \boldsymbol{\nabla}_\mathbf{x} \mathbf{u} + \boldsymbol{\nabla}_\mathbf{x} p = \nu \triangle_\mathbf{x} \mathbf{u} \tag{1}$$

for the Eulerian velocities $\mathbf{u}(\mathbf{x}, t)$ and the kinematic pressure $p(\mathbf{x}, t)$ of incompressible motion of a fluid with kinematic viscosity $\nu$ in a three-dimensional flow volume $\Omega$.

To be more definite, we consider plane shear flow between two infinite plates (Fig. 1). We define the plates to be located at $x_2 = 0$ moving with velocity $U_L$ in positive $x_3$-direction respectively resting at $x_2 = L$. Reynolds' number is defined as $Re = U_L L/\nu$.

### A. Velocity increments

The choice of the order parameter is crucial for any RG theory. It is therefore desirable to find a closed equation of motion for velocity increments instead of considering the Navier–Stokes equation for the Eulerian velocities, since such variables are commonly believed to be capable of exhibiting universal behavior. This can be achieved by means of a transformation introduced by Monin [57,58]. Since it does not seem to be widely known in spite of its reintroduction in [53], we present it here relatively detailed. The idea is to measure the Eulerian velocity field $\mathbf{u}$ in a frame of reference defined by an arbitrary but fixed marker particle passively advected with the flow (Fig. 2). The Lagrangian trajectory $\mathbf{s}(t|\mathbf{x}_0, t_0)$ for such a marker which is at position $\mathbf{x} = \mathbf{x}_0$ for $t = t_0$ is given by

$$\mathbf{s}(t|\mathbf{x}_0, t_0) = \mathbf{x}_0 + \int_{t_0}^{t} d\tau \, \mathbf{u}(\mathbf{s}(\tau|\mathbf{x}_0, t_0), \tau). \tag{2}$$

Next, we define a "velocity increment"

$$\mathbf{w}(\mathbf{r}, t|\mathbf{x}_0, t_0) = \mathbf{u}(\mathbf{s}(t|\mathbf{x}_0, t_0) + \mathbf{r}, t) - \mathbf{u}(\mathbf{s}(t|\mathbf{x}_0, t_0), t). \tag{3}$$

The field $\mathbf{w}(\mathbf{r}, t)$ describes a velocity difference between two space points separated by the increment $\mathbf{r}$, for all times $t$, but has Eulerian as well as Lagrangian characteristics. Monin's transformation from $\mathbf{u}$ to $\mathbf{w}$ thus takes care of the very special property of flow fields being coupled to the actual movement of fluid elements.

Plugging (3) into (1) gives a closed equation for $\mathbf{w}$,

$$\partial_t \mathbf{w} + \mathbf{w} \cdot \boldsymbol{\nabla}_\mathbf{r} \mathbf{w} + \boldsymbol{\nabla}_\mathbf{x} p = \nu \triangle_\mathbf{r} \mathbf{w} - \ddot{\mathbf{s}}. \tag{4}$$

There is no explicitly $\mathbf{u}$ dependent term due to Galilean invariance. The term $-\ddot{\mathbf{s}}$ describes a fictitious force (per mass) because the frame of reference, i.e., the marker, moves nonuniformly. Since it is independent of $\mathbf{r}$, it may be eliminated by taking the difference of (4) for arbitrary $\mathbf{r}$ as well as for $\mathbf{r} = 0$. Formally, we introduce an operator $\mathcal{S}$,

$$\mathcal{S}[\mathbf{q}](\mathbf{r}, t) = \mathbf{q}(\mathbf{r}, t) - \mathbf{q}(0, t), \tag{5}$$

and apply it to (4):

$$\partial_t \mathbf{w} + \mathbf{w} \cdot \boldsymbol{\nabla}_\mathbf{r} \mathbf{w} + \mathcal{S}[\boldsymbol{\nabla}_\mathbf{x} p] = \nu \, \mathcal{S}[\triangle_\mathbf{r} \mathbf{w}]. \tag{6}$$

The boundary conditions cannot be expressed in terms of $\mathbf{w}$ and $\mathbf{r}$ alone since the boundaries brake Galilean invariance. The velocity increments $\mathbf{w}$ are defined on a flow volume moving relative to the marker,

$$\Omega(\mathbf{s}(t|\mathbf{x}_0, t_0)) = \{\mathbf{r} = \mathbf{x} - \mathbf{s}(t|\mathbf{x}_0, t_0) | \mathbf{x} \in \Omega\}$$

with the corresponding surface $\partial\Omega(\mathbf{s}(t|\mathbf{x}_0, t_0))$. So far our introduction of Monin's transformation.

We now specify how statistical averages are defined. First, a local $\mathbf{x}$-centered space average over an $a$-cube is taken for every space point $\mathbf{x}$. Second, an infinite time average is performed. Finally, the limit of the averaging volume approaching zero is taken. We assume that the properties of the nonlinear fluid dynamics guarantee that all mean quantities considered in this work exist and are independent of $x_1$, $x_3$ and $t_0$. More formally, we define:

$$\begin{aligned}
\langle f \rangle (x_2) &= \lim_{a \to 0} \lim_{t \to \infty} \int_{t_0}^{t} \frac{d\tau}{t - t_0} \int_{2|\xi_i - x_i| \leq a} \frac{d^3\boldsymbol{\xi}}{a^3} f(\boldsymbol{\xi}, \tau) \\
&= \lim_{a \to 0} \lim_{t \to \infty} \int_{t_0}^{t} \frac{d\tau}{t - t_0} \int_{2|s_i(\tau|\mathbf{x}_0, t_0) - x_i| \leq a} \frac{d^3\mathbf{x}_0}{a^3} f(\mathbf{r}, \tau|\mathbf{x}_0, t_0).
\end{aligned} \quad (7)$$

The second equality follows from $\det \boldsymbol{\nabla}_{\mathbf{x}_0}\mathbf{s} = 1$ which holds because $\boldsymbol{\nabla}_{\mathbf{s}}\cdot\mathbf{u} = 0$ and Lagrangian trajectories are unique.

From now on we will often use the shorthand notation $\mathbf{s}(t)$ and $\mathbf{w}(\mathbf{r}, t)$ for $\mathbf{s}(t|\mathbf{x}_0, t_0)$ and $\mathbf{w}(\mathbf{r}, t|\mathbf{x}_0, t_0)$, respectively.

### B. $r$-velocity fluctuations

The statistics of the velocity increment field $\mathbf{w}$ are nonuniversal in general. Not even the mean value $\langle \mathbf{w} \rangle = \mathbf{U}(x_2 + r_2) - \mathbf{U}(x_2)$ will vanish because of the anisotropy inherent to the turbulent profile $U_i(x_2) = \langle u_i \rangle (x_2) = U(x_2)\,\delta_{i,3}$ which reflects carries the boundary anisotropy. It seems therefore reasonable to subtract the turbulent profile from both Eulerian velocities appearing in the definition of $\mathbf{w}$. We call the resulting quantity

$$\begin{aligned}
\mathbf{v}(\mathbf{r}, t) &= \mathbf{w}(\mathbf{r}, t) - [\mathbf{U}(\mathbf{s}(t) + \mathbf{r}) - \mathbf{U}(\mathbf{s}(t))] \\
&= [\mathbf{u}(\mathbf{s}(t) + \mathbf{r}, t) - \mathbf{U}(\mathbf{s}(t) + \mathbf{r})] - [\mathbf{u}(\mathbf{s}(t), t) - \mathbf{U}(\mathbf{s}(t))]
\end{aligned} \quad (8)$$

the "$r$-velocity fluctuation". By definition we have $\langle \mathbf{v} \rangle = 0$ and, of course, $\mathbf{v}$ is divergence free. The boundary condition for $\mathbf{v}(\mathbf{r}, t)$ is given by the marker's negative $r$-velocity fluctuations,

$$\mathbf{v}(\mathbf{r},t)\big|_{\mathbf{r}\in\partial\Omega(\mathbf{s}(t))} = -\left[\mathbf{u}(\mathbf{s}(t),t) - \mathbf{U}(\mathbf{s}(t))\right], \tag{9}$$

which, remarkably, is independent of $\mathbf{r}$. The quantity $\mathbf{v}$ is expected to have universal properties and seems to be the most natural choice for the "order parameter" of turbulence. It vanishes for laminar flows.

The pressure $p$ is eliminated as usual by formally solving a Poisson equation (we employ standard summation conventions):

$$p(\mathbf{x},t) = p_\mathrm{f}(\mathbf{x},t) - \int_\Omega d^3\boldsymbol{\xi}\, G_\Omega(\mathbf{x},\boldsymbol{\xi})\, \partial_{\xi_i}\partial_{\xi_j}\big(u_i(\boldsymbol{\xi},t)u_j(\boldsymbol{\xi},t)\big),$$

where the Green function $G_\Omega$ is determined by the geometry, i.e., $\partial_{x_i}\partial_{x_i}G_\Omega(\mathbf{x},\boldsymbol{\xi}) = \delta(\mathbf{x}-\boldsymbol{\xi})$, $G_\Omega(\mathbf{x},\boldsymbol{\xi})\big|_{\mathbf{x}\in\partial\Omega} \equiv 0$. Here, the pressure is assumed to be constant at the closed parts of the boundary, otherwise there would be an extra surface contribution to $p$. While this is physically reasonable, it should be remarked that the admissibility of such boundary conditions is mathematically nontrivial, since the compatibility condition $\nu\triangle_\mathbf{x}\mathbf{u} = \boldsymbol{\nabla}_\mathbf{x} p$ has to be fulfilled also at the boundary. For the geometry of plane shear flow one can easily verify the integral representation

$$G_\Omega(\mathbf{x},\boldsymbol{\xi}) = -\frac{1}{4\pi}\int_0^\infty dk\, J_0(k\sqrt{(x_1-\xi_1)^2 + (x_3-\xi_3)^2})\cdot$$
$$\cdot\frac{\cosh(k(|x_2-\xi_2|-L)) - \cosh(k(x_2+\xi_2-L))}{\sinh(kL)} \tag{10}$$

for the Green function. The expression for the pressure may be combined with the nonlinear term of the Navier–Stokes equation,

$$\mathbf{u}\cdot\boldsymbol{\nabla}_\mathbf{x}\mathbf{u} + \boldsymbol{\nabla}_\mathbf{x}p = \widetilde{\mathcal{D}}[\mathbf{u}\cdot\boldsymbol{\nabla}_\mathbf{x}\mathbf{u}] + \boldsymbol{\nabla}_\mathbf{x}p_\mathrm{f},$$

where the operator $\widetilde{\mathcal{D}}$ projects to divergence free functions:

$$\widetilde{\mathcal{D}}_i[\mathbf{q}](\mathbf{x},t) = q_i(\mathbf{x},t) - \int_\Omega d^3\boldsymbol{\xi}\, \partial_{x_i}G_\Omega(\mathbf{x},\boldsymbol{\xi})\, \partial_{\xi_j}q_j(\boldsymbol{\xi},t).$$

Finally, we obtain an equation of motion for $\mathbf{v}$ by inserting (8) into (6) and employing $\boldsymbol{\nabla}_{\mathbf{r}}\mathbf{w} = \boldsymbol{\nabla}_{\mathbf{r}}\mathbf{v} + \boldsymbol{\nabla}_{\mathbf{r}}\mathbf{U}(\mathbf{s}+\mathbf{r})$:

$$\partial_t \mathbf{v} + \mathcal{SD}[\mathbf{v}\cdot\boldsymbol{\nabla}_{\mathbf{r}}\mathbf{v}] = \nu\,\mathcal{S}[\triangle_{\mathbf{r}}\mathbf{v}] + \mathcal{F}[\mathbf{v}]. \tag{11}$$

The projector $\mathcal{D}$ is defined analogously to $\widetilde{\mathcal{D}}$:

$$\mathcal{D}_i[\mathbf{q}](\mathbf{r},t) = q_i(\mathbf{r},t) - \int_{\Omega(\mathbf{s}(t))} d^3\boldsymbol{\rho}\, \partial_{r_i} G_\Omega(\mathbf{r}+\mathbf{s}(t), \boldsymbol{\rho}+\mathbf{s}(t))\, \partial_{\rho_j} q_j(\boldsymbol{\rho},t). \tag{12}$$

The terms containing the turbulent profile $\mathbf{U}$ can be interpreted as driving force (per mass) for the $r$-velocity fluctuation:

$$\begin{aligned}\mathcal{F}[\mathbf{v}] = &-\mathcal{SD}[(\mathbf{U}(\mathbf{s}+\mathbf{r}) - \mathbf{U}(\mathbf{s}))\cdot\boldsymbol{\nabla}_{\mathbf{r}}\mathbf{v}(\mathbf{r},t) + \mathbf{v}(\mathbf{r},t)\cdot\boldsymbol{\nabla}_{\mathbf{r}}\mathbf{U}(\mathbf{s}+\mathbf{r})] \\ &+ \nu\,\mathcal{S}[\triangle_{\mathbf{r}}\mathbf{U}(\mathbf{s}+\mathbf{r})] - \mathcal{S}\boldsymbol{\nabla}_{\mathbf{r}} p_{\mathrm{f}} - \mathbf{u}(\mathbf{s},t)\cdot\mathcal{S}[\boldsymbol{\nabla}_{\mathbf{r}}(\mathbf{U}(\mathbf{s}+\mathbf{r}) - \mathbf{U}(\mathbf{s}))].\end{aligned} \tag{13}$$

The forcing term depending explicitly on $\mathbf{u}$ is equal to $\partial_t[\mathbf{U}(\mathbf{s}(t)+\mathbf{r}) - \mathbf{U}(\mathbf{s}(t))]$. Such explicit $\mathbf{u}$ dependence is no surprise since the Galilean invariance, which Monin's transformation relies on, is destroyed by the boundary. Thus, (11) alone is not equivalent to (1) because the trajectory $\mathbf{s}(t)$ is needed which itself is determined by (1).

We consider (11) as an equation of motion for the order parameter $\mathbf{v}$ with given profile $\mathbf{U}$ and marker trajectories $\mathbf{s}(t)$. In this sense it is analogous to the various time-dependent Ginzburg–Landau equations used for RG treatment of dynamical critical phenomena, but as opposed to the latter the forcing here is deterministic. No statistical distribution has to be introduced as it is usually done for RG theories of turbulence.

### III. A NEW RG METHOD

In this section we outline our physical understanding how a RG theory could be applied to fully developed turbulence. The point of view presented in Sect. III A will guide us in Sect. III B where the ideas for a new RG method are developed.

## A. Inverse RG

After choosing the $r$-velocity fluctuation $\mathbf{v}$ as the order parameter in the previous section, a mapping of quantities of turbulence to those of critical phenomena is given. It is based on the idea of an inverse RG transformation pursued in earlier work [46–50].

The large spatial scales, which are governed by a forcing depending on the flow geometry, are considered as nonuniversal in the RG sense, and thus correspond to the microscopic interaction scales (or lattice constant) in critical phenomena. The small spatial scales where viscosity dominates are seen as universal in the RG sense and analogous to the scales in critical phenomena set by the macroscopic system size. In this picture one expects anomalous scaling for $r$ above $\ell$ in analogy to $r$ below the correlation length in the critical case and regular behaviour for $r \to 0$ as for the wavenumber approaching zero in the critical case. Thus, the (longitudinal) structure functions

$$D^{(m)}(\mathbf{r}) = \left\langle \left(\mathbf{v}(\mathbf{r},t) \cdot \frac{\mathbf{r}}{r}\right)^m \right\rangle \tag{14}$$

should scale with non-K41 intermittency exponents in the inertial range, and be analytic in the viscous range. We do not study correlations in time and implicitly allege that turbulence has a "static limit", i.e., a description of equal-time correlations independent of correlations in time.

At this point one might ask which turbulent quantity corresponds to the correlation length. If we take the exchange of the role of large and small scales serious, this ought to be a viscous length which we will call $\ell$ in the following. Since there is strong indication that the energy spectrum decays in the viscous range according to the universal law

$$E(k) \propto \left\langle |\mathbf{v}(\mathbf{k},t)|^2 \right\rangle \propto \exp(-\ell\, k/(2\pi))\, k^\beta \quad \text{for} \quad 2\pi/k \ll \ell$$

with some exponent $\beta$ [32], we may define a dissipation length $\ell(\nu)$ through $\ell = -\lim_{k\to\infty}(\ln E(k))/k$. This is completely analogous to the definition of the correlation length through the decay of the order parameter correlation function for spatial distances far beyond the correlation length.

Because there is no evidence that $\ell$ vanishes for a finite value of $\nu$, $\nu \to 0$ is the "turbulent limit" corresponding to the critical limit of the temperature approaching the critical temperature. Using the language of phase transitions, a turbulent system is always in the "disordered phase" because of $\langle \mathbf{v} \rangle = 0$. Since an ordered phase would appear for negative, but physically forbidden $\nu$ only, we do not expect a phenomenon analogous to spontaneous symmetry breaking. As the critical limit is defined for fixed microscopic parameters, we define the turbulent limit for fixed macroscopic quantities, i.e., fixed geometry and outer scale $L$ as well as fixed outer velocity $U_L$. This is the precise definition of what we understand as "$Re \to \infty$". Thus, $\nu$ is supposed to be the only critical parameter as opposed to temperature and external magnetic field in critical phenomena; $\mathbf{U}$ is analogous to the coupling parameter.

Recognizing this analogy to critical phenomena, the first step of a RG transformation for turbulence should eliminate degrees of freedom on the macroscopic scale $L$, resulting in a new field $\hat{\mathbf{U}}$ which is defined in a *smaller* system with size $\hat{L} = L/b$ where $b > 1$ is the usual RG rescaling parameter. In the second step the system is rescaled with $b$ to its original size $L' = b\hat{L} = L$ and the equation of motion should take on the same *functional* form with renormalized turbulent velocity fluctuation $\mathbf{v}'$, viscosity $\nu'$, and turbulent profile $\mathbf{U}'$, which is analogous to the renormalization of the order parameter, the temperature, and the coupling parameters in critical phenomena. $\mathbf{U}$ should approach a fixed point under renormalization. The spatial rescaling implies an increase by a factor of $b$ of all *physical* microscopic length scales, in particular $\ell' = b\ell$ for the dissipation length.

Coming back to the central question of intermittency corrections, we give the functional form structure functions are expected to show for $\nu \to 0$:

$$D^{(m)}(\mathbf{r}, \nu, L) = \begin{cases} b_m \left(\varepsilon^\star r\right)^{m/3} \left(\frac{r}{L}\right)^{\delta \zeta_m} & \text{for} \quad \ell \ll r \ll L \\ D_0^{(m)}(\nu, L) \, r^m & \text{for} \quad r \ll \ell \ll L \end{cases}. \quad (15)$$

Here, the energy dissipation rate $\varepsilon^\star$ is supposed to be finite for $\nu \to 0$; this will be defined and discussed later. The constants $b_m$ are also supposed to be finite for $\nu \to 0$. Thus, intermittency corrections are connected to the outer length scale $L$, whereas in critical phenomena the critical exponents' deviations from their Gaussian values show up as a dependence of correlation function amplitudes on the microscopic scale.

The form (15) is also compatible with the concavity of the curve $\delta\zeta_m$ as a function of $m$, as a simple application of Schwarz' inequality applied to $\langle v^m \rangle$ shows, leading to $2\,\delta\zeta_{m_1+m_2} \geq \delta\zeta_{2m_1} + \delta\zeta_{2m_2}$ if $L$ defines the correction scale, so $r/L \leq 1$ [33,59]. But this argument does not rule out more complicated scenarios such as the simultaneous occurrence of $L$ and $\ell$ as correction scales. Also, we have assumed that structure functions are isotropic in the inertial and viscous ranges despite of an anisotropic geometry and that they are independent of $x_2$ at least near the centre of the flow $x_2 = L/2$ where we expect universal behaviour.

In the preceding section we recognized another macroscopic length besides $L$ entering the equation of motion: the marker trajectories $\mathbf{s}(t)$, for which there does not seem to be an analogue in critical phenomena. To be more precise, only the $s_2$ component of $\mathbf{s}$ is relevant because of translational invariance in planes parallel to the plates. We refer to $s_2(t)$ as a "dynamic outer length" as opposed to $L$ being a "geometric outer length". Because of its time dependence it cannot function as a correction scale in (15), but it has to be dealt with during renormalization, cf. Sect. IV.

We consider such an inverse RG to make more sense physically than a formal application of RG procedures for critical phenomena to turbulence. But this insight does not tell us how a procedure for eliminating degrees of freedom or a field-theoretic RG theory should be constructed. The main problem is our missing information about the statistical distribution for the $r$-velocity fluctuation.

### B. A new idea

*Instead of trying to construct a specific RG procedure based on some assumed statistical distribution, the new idea of this work is to make a general ansatz for possible RG transformations and to examine their scope under physically motivated constraints.*

Instead of specifying a procedure for eliminating degrees of freedom, we consider an "ar-bitrary" transformation $\mathbf{v}(\mathbf{r}, t) \mapsto \mathbf{v}'(\mathbf{r}', t')$ of the $r$-velocity fluctuation. But we demand that the renormalized $r$-velocity fluctuation $\mathbf{v}'$ obeys "essentially" the same equation of motion. This will reduce the number of transformations admitted but, in addition, we have to pose sufficiently many, physically motivated constraints which reduce the number of transformations admitted even further such that nontrivial statements about scaling exponents become possible. Behind this idea is the hope that only a limited amount of RG transformations is compatible with the Navier–Stokes equation.

The most important constraint to the equation of motion for $\mathbf{v}'$ is the invariance of the nonlinear term since the structure of the nonlinearity is prescribed by hydrodynamics and must not change. To give an example, because of Galilean invariance $\mathbf{u} \cdot \boldsymbol{\nabla} \mathbf{u}$ must not be multiplied by a coupling parameter which changes under RG transformations. Although the $r$-velocity fluctuation $\mathbf{v}$ itself is Galilean invariant, such a variable coupling parameter would contradict the Navier–Stokes equation from which the equation of motion for $\mathbf{v}$ results.

Since the markers' Eulerian velocity $\mathbf{u}(\mathbf{s}, t)$ enters the equation of motion for $\mathbf{v}$, the transformation of this quantity has to be specified, too. We demand that the equation of motion for $\mathbf{u}'(\mathbf{s}', t')$ shall again be a Navier–Stokes equation which might differ from that for $\mathbf{u}$ only in the dissipative term. This is even more restricting than what we demand for the transformation of $\mathbf{v}$. But we allow that $\mathbf{v}'$ does not have to be a difference of Eulerian velocities $\mathbf{u}'$ according to (8), because otherwise the transformation of $\mathbf{v}$ would be fully determined by that of $\mathbf{u}$, and there would be no need and no gain to introduce $\mathbf{v}$ in the beginning. Since we believe velocity differences to be of fundamental importance, we look at the variables $\mathbf{v}$ and $\mathbf{u}$ as independent velocity fields which might "decouple" in the course of renormalization. Here, we do not see any analogy to critical phenomena. Note that the representation of the equation of motion (11) in terms of $\mathbf{v}$, $\mathbf{u}$, and $\mathbf{U}$ is unique if only Eulerian velocities at the current marker position $\mathbf{s}(t)$ enter. Although the reasoning for the invariance of the nonlinearity is based on the connection between the equation of motion for $\mathbf{v}$ and the Navier–Stokes equation, we keep this demand also if $\mathbf{v}$ and $\mathbf{u}$ decouple as outlined above.

The second big constraint comes from the assumed special role of energy dissipation. It seems natural to assign the central role which the free energy plays in RG theory of critical phenomena to the energy dissipation rate in turbulence [48,51].

The energy dissipation rate (per mass) averaged over the full flow volume $\Omega$,

$$\varepsilon_\Omega = \frac{\nu}{2} \lim_{t \to \infty} \int_{t_0}^{t} \frac{d\tau}{t - t_0} \int_\Omega \frac{d^3\mathbf{x}}{\Omega} \Big( \partial_{x_j} u_i(\mathbf{x}, \tau) + \partial_{x_i} u_j(\mathbf{x}, \tau) \Big)^2 \quad (16)$$

is a fundamental physical quantity in turbulence since it equals the energy injection rate $\varepsilon_{\text{in}}$ to maintain dynamical stationarity. We see it as the analogue to the free energy density in critical phenomena and expect it to be finite in the turbulent limit. This corresponds to the widespread belief that the limit $\nu \to 0$ is fundamentally different from the Eulerian case $\nu = 0$ where $\varepsilon_\Omega = 0$.

In RG theory of critical phenomena it is not the free energy *density* itself which is RG invariant but the *total* free energy. For turbulence, we suggest to also make such a distinction. Since $\varepsilon_{\text{in}} = \nu U_L (\partial_{x_2} U(x_2))|_{x_2=0}/L$ (for plane shear flow) is determined by the mostly nonuniversal behaviour of the turbulent profile at the boundaries, $\varepsilon_\Omega$ does not seem to be well suited as a RG invariant quantity. Instead we choose the turbulent energy dissipation rate $\varepsilon_{\text{t}}$, which lies at the heart of Kolmogorov-based phenomenology, as an analogue to the free energy. To define it, we start from the mean local dissipation rate

$$\varepsilon(x_2) = \frac{\nu}{2} \left\langle \Big( \partial_{x_j} u_i(\mathbf{x}, t) + \partial_{x_i} u_j(\mathbf{x}, t) \Big)^2 \right\rangle \quad (17)$$

near the centre of fluid motion ($x_2 \approx L/2$). The averaging procedure was defined in (7).

We now rewrite $\varepsilon$ as a functional of $r$-velocity fluctuations and split off the dissipation related directly to the turbulent profile: $\varepsilon = \varepsilon_{\text{U}} + \varepsilon_{\text{t}}$ with

$$\begin{aligned} \varepsilon_{\text{U}}(x_2) &= \frac{\nu}{2} \Big( \partial_{x_2} U(x_2) \Big)^2, \\ \varepsilon_{\text{t}}(x_2) &= \frac{\nu}{2} \left\langle \Big( \partial_{r_j} v_i(\mathbf{r}, t) + \partial_{r_i} v_j(\mathbf{r}, t) \Big)^2 \right\rangle \Big|_{\mathbf{r}=0}. \end{aligned} \quad (18)$$

This splitting is unique since the mixed term involving both $\mathbf{U}$ and $\mathbf{v}$ vanishes. The profile's dissipation $\varepsilon_{\text{U}}$ is nonuniversal and corresponds to the background term present in the free energy known from RG theory. But we do not know if the quantity $\int_0^L \frac{dx_2}{L} \varepsilon_U(x_2)$ is analytic in $\nu$ as its correspondence to the background term in the free energy density, being analytic in the temperature, demands. We remark that $\varepsilon_t$ equals the energy dissipation rate coming from the fluctuations $\mathbf{u} - \mathbf{U}$, i.e.,

$$\varepsilon_t(x_2) = \frac{\nu}{2} \left\langle \left( \partial_{x_j} \big(u_i(\mathbf{x},t) - U_i(\mathbf{x})\big) + \partial_{x_i} \big(u_j(\mathbf{x},t) - U_j(\mathbf{x})\big) \right)^2 \right\rangle.$$

This turbulent energy dissipation rate $\varepsilon_t$ shall be our "RG invariant quantity". This means two things, again analogous to RG theory for critical phenomena: first, any RG transformation should preserve the functional structure of $\varepsilon_t$ when written as a functional of $\mathbf{v}(\mathbf{r},t)$ with parameters $\nu$, $x_2$, and $L$ — except for an additive, $\mathbf{v}$ independent contribution. This corresponds to the functional invariance of the free energy density; again up to an additive, order parameter independent contribution. The main difference to critical phenomena is the missing knowledge about the probability distribution functional for $\mathbf{v}(\mathbf{r},t)$. Instead, we work with the purely formal expression (18) for $\varepsilon_t$. Note that this might not be the same as using the definition (17) since $\mathbf{v}$ and $\mathbf{u}$ might be transformed differently. Second, and corresponding to the invariance of the free energy, the additive constant is put into $\varepsilon'_U$ such that the renormalized value $\varepsilon'$ of the local dissipation rate equals the original value $\varepsilon$:

$$\varepsilon' = \varepsilon'_U + \varepsilon'_t = \varepsilon_U + \varepsilon_t = \varepsilon. \tag{19}$$

The aforementioned $\varepsilon^\star$ is now defined to be the turbulent limit of $\varepsilon_t$, i.e. $\varepsilon^\star = \varepsilon_t|_{\nu \to 0}$ and is also assumed to be finite for $\nu \to 0$. For K41 scaling behaviour we identify the Kolmogorov length $\eta = (\nu^3/\varepsilon^\star)^{1/4}$ with the dissipation length $\ell$, up to a dimensionless factor which is assumed to be a finite constant for $\nu \to 0$. If there are intermittency corrections, however, $\ell$ should depend in a nontrivial way on $\nu$, just as the correlation length scales nontrivially with the temperature parameter. On the other hand, the RG transformation behaviour should be $\ell' = b\ell$ (analogous to a decrease of the critical correlation length by a factor $b$), whereas in general $\eta' \neq b\eta$ if $\nu$ scales nontrivially.

The other constraints are of minor importance and will be mentioned in the course of fully specifying our RG method in the following sections.

But before going into the details we can already infer some of the advantages and disadvantages of such an RG approach. Since the forcing enters only structurally in the equation of motion for **v**, and we do not actually eliminate degrees of freedom, we cannot expect to be able to really calculate non-trivial scaling exponents. This is only achievable for critical phenomena since much more knowledge about the forcing, coming from thermodynamics, is made good use of. But we hope to learn if and how the lack of a boundary enforces the classical exponents and if there are universal relations between exponents. In our approach, we assume the validity of the "inverse RG" analogy to critical phenomena. The existence of RG transformations with the desired properties including consequences of symmetries is assumed, not shown by construction. Despite these disadvantages, we overcome several shortcomings of previous RG methods in turbulence. We do not have to know the probability distribution for some velocity field or make wild assumptions about it. Our method is nonperturbative which is nice since fully developed turbulence has no obvious expansion parameter.

## IV. RG TRANSFORMATION

We come to the actual presentation of our RG method for turbulence, following the spirit of the previous section. Because of the invariance of the nonlinearity the original equation of motion and its renormalized counterpart should have the following form (primes denote renormalized quantities):

$$\partial_t \mathbf{v} + \mathcal{SD}[\mathbf{v} \cdot \boldsymbol{\nabla}_\mathbf{r} \mathbf{v}] = \mathcal{L}[\mathbf{v}] \equiv \nu \, \mathcal{S}[\triangle_\mathbf{r} \mathbf{v}] + \mathcal{F}[\mathbf{v}], \tag{20}$$

$$\partial_{t'} \mathbf{v}' + \mathcal{SD}'[\mathbf{v}' \cdot \boldsymbol{\nabla}_{\mathbf{r}'} \mathbf{v}'] = \mathcal{L}'[\mathbf{v}'] \equiv \nu' \, \mathcal{S}[\triangle_{\mathbf{r}'} \mathbf{v}'] + \mathcal{F}'[\mathbf{v}']. \tag{21}$$

Here, the terms linear in **v**, respectively **v**′, are combined by introducing an inhomogeneous linear operator $\mathcal{L}$, respectively $\mathcal{L}'$. Renormalization of forcing and dissipation expresses itself as $\mathcal{L}' \neq \mathcal{L}$. Besides $\nu' > \nu$, one expects to find structurally new terms, which we initially consider as part of $\mathcal{F}' \neq \mathcal{F}$. They have to be interpreted as 'dissipative' or 'driving' on an individual basis.

In order for equation (21) to make sense, the transformation of $\mathbf{r}$, $t$, $\mathbf{v}$, $\mathbf{s}(t)$, and $\mathbf{u}$ has to be defined. Finding transformations such that (21) has the prescribed structure is the main goal of Sects. IV and V. We begin with discussing the parameters which these transformations may depend on.

An important assumption is the regularity of RG transformations for $\nu \to 0$. For critical phenomena, the analogous regularity in the temperature approaching the critical temperature is a main achievement of RG theory. In this work we concentrate on the universal scaling behaviour in the inertial range and thus may already carry out the limit $\nu \to 0$ *in the transformation*. Effectively, it suffices to take $\nu$ independent transformations. As long as we do not look at multiple-time correlations of $\mathbf{v}$, we assume that the transformations do not have to be $t$ dependent. But $\mathbf{r}$ should be admitted, as well as both outer lengths $L$ and $s_2(t)$ since all these are nonuniversal. Dismissing $s_1$ and $s_3$ takes into account geometric symmetries, and the possible dependence on $s_2(t)$ allows an implicit time dependence. We even start with the admission of all components of $\mathbf{s}$ in order to study the consequences of the restriction to $s_2$ later. Note, however, that we will not enforce this restriction for the transformation of $\mathbf{u}$; it will be allowed to depend on all components of $\mathbf{s}$. With the prelude in mind, we come to the specification of our ansatz for RG transformations.

### A. Space and time

Spatial distances are just rescaled:

$$\mathbf{r}' = (1 + \sigma)\mathbf{r} + \mathcal{O}(\sigma^2), \quad \text{thus} \quad \partial_{r_i'} = (1 - \sigma)\partial_{r_i} + \mathcal{O}(\sigma^2). \tag{22}$$

The RG rescaling factor $b = 1 + \sigma$ deviates only infinitesimally from 1 by the RG flow parameter $\sigma$. Experience with critical phenomena teaches that such infinitesimal RG transformations usually exist and are even easier to deal with. Because the RG transformation shall be inverse, we have $r' > r$. For time differences we will also begin with a simple rescaling ansatz,

$$t' = t + \sigma\,(t - t_0)\,z + \mathcal{O}(\sigma^2), \tag{23}$$

with a single constant scaling exponent $z$. In a supplement to Sect. V A we will examine a more general transformation of $\mathbf{r}$ and spatially varying $z$.

### B. Marker trajectories

Because of the presence of the marker trajectories $\mathbf{s}(t|\mathbf{x}_0, t_0)$ in the equation of motion we have to specify how they shall transform in spite of not knowing a corresponding quantity in critical phenomena. Since the trajectories are nonuniversal their transformation could and should be different from the transformation (22) of distances $\mathbf{r}$.

Because the trajectory's position $\mathbf{s}(t)$ enters the argument of the Green function in (12), its transformation is linked to that of the geometry. We demand that the geometry, and thus the Green function, is RG invariant. This corresponds to keeping the geometric structure of the microscopic lattice fixed in critical phenomena. On the other hand, the RG transformation should change the nature of the forcing as it changes the microscopic interaction for critical phenomena. Thus, keeping the geometry invariant actually assumes that all renormalization of the forcing can be covered by a change of the linear forcing $\mathcal{F}$. Because of this invariance of geometry and $L' = L$ we have to define $\mathcal{D}'$ in (20) to be

$$\mathcal{D}'_i[\mathbf{q}](\mathbf{r}, t) = q_i(\mathbf{r}, t) - \int_{\Omega(\mathbf{s}'(t'))} d^3\boldsymbol{\rho}'\,\partial_{r'_i} G_\Omega(\mathbf{r}' + \mathbf{s}'(t'), \boldsymbol{\rho}' + \mathbf{s}'(t'))\,\partial_{\rho'_j} q_j(\boldsymbol{\rho}, t). \tag{24}$$

The most simple way to satisfy this geometric invariance is to keep the marker's position the same, i.e., $\mathbf{s}'(t') = \mathbf{s}(t)$. For a fixed fluid element serving as the marker this is impossible since the dynamics and the trajectory of a fixed marker change inevitably under RG transformations. But we may allow the marking fluid element to change. We choose, for every time $t$, a marker with initial position $\mathbf{x}'_0$ such that

$$\mathbf{s}(t) \equiv \mathbf{s}(t|\mathbf{x}_0, t_0) = \mathbf{s}(t'|\mathbf{x}'_0, t_0) \equiv \mathbf{s}'(t'). \tag{25}$$

This definition is unique because of the one-to-one correspondence of initial positions $\mathbf{x}_0$ and current positions $\mathbf{s}(t|\mathbf{x}_0, t_0)$ for all $t$. It also guarantees $\mathbf{s}'(t') \in \Omega$. The set of all transformed trajectories $\mathbf{s}'$ naturally defines an Eulerian velocity field

$$\mathbf{u}'(\mathbf{s}', t') = \dot{\mathbf{s}}' \equiv \partial_{t'}\mathbf{s}'(t'|\mathbf{x}'_0, t_0)\big|_{\mathbf{x}'_0} \tag{26}$$

which should satisfy a Navier–Stokes equation. We will cover the transformation of $\mathbf{u}$ in Sect. VI B.

Going from $\mathbf{x}_0$ to $\mathbf{x}'_0$ enforces to also relate the transformed fluctuation $\mathbf{v}'$ to a marker with initial position $\mathbf{x}'_0$ instead of $\mathbf{x}_0$. Therefore the time derivative $\partial_{t'}\mathbf{v}'$ in (21) has to be taken with fixed $\mathbf{x}'_0$. From (23), (25) and (26) we then find

$$\begin{aligned}\partial_{t'} \equiv \partial_{t'}\big|_{\mathbf{r}',\mathbf{x}'_0} &= (1 - \sigma z)\, \partial_t\big|_{\mathbf{r},\mathbf{x}_0} + [\mathbf{u}' - (1 - \sigma z)\,\mathbf{u}]\cdot\boldsymbol{\nabla}_{\mathbf{s}}\big|_{\mathbf{r},t} + \mathcal{O}(\sigma^2) \\ &= (1 - \sigma z)\, \partial_t\big|_{\mathbf{s},\mathbf{r}} + \mathbf{u}'\cdot\boldsymbol{\nabla}_{\mathbf{s}}\big|_{\mathbf{r},t} + \mathcal{O}(\sigma^2).\end{aligned} \tag{27}$$

### C. Statistical averages

Mean values $\langle f \rangle'$ are defined analogously to (7) with the corresponding transformed quantities. For the transformations defined so far we simply have $\langle f \rangle' = \langle f \rangle$. To see this, we only have to look at the spatial averaging:

$$\int_{2|\mathbf{s}'_i(\tau') - x_i| \le a'} \frac{d^3\mathbf{x}'_0}{a'^3} = \int_{2|\mathbf{s}_i(\tau) - x_i| \le a'} \frac{d^3\mathbf{x}'_0}{a'^3}$$
$$= \int_{2|\mathbf{s}_i(\tau) - x_i| \le a} \frac{d^3\mathbf{x}_0}{a^3}\left[1 - 3\sigma + \sigma\frac{a}{2}\sum_j \left[\delta(s_j - x_{0;j} - \tfrac{a}{2}) + \delta(s_j - x_{0;j} + \tfrac{a}{2})\right] + \mathcal{O}(\sigma^2)\right].$$

Here we made use of (25) and $a' = (1 + \sigma)a + \mathcal{O}(\sigma^2)$. The latter holds because $a$ is a microscopic length which should be rescaled within an inverse RG theory with a factor $b$ (with no nontrivial exponent, cf. Sect. III A) — independent of possible intermittency corrections because we identify this formal parameter $a$ with a physical quantity, namely the linear size of fluid elements, which we also designate as the hydrodynamic length scale. The formal limit $a \to 0$ corresponds to $a \ll \ell$, i.e., one is only interested in the hydrodynamic length being smaller than the dissipative scale. This role of $a$ is analogous to that of the system size $L$ in critical phenomena, where the case $L$ exceeding the correlation length is of main interest. Just as finite size effects do not change the critical exponents, we expect intermittency exponents to be independent of effects on scales less than $a$.

We now assume that the surface averages represented by the terms with a delta function equal the full average $\langle\rangle$, at least after taking the time average and limits according to (7). Then all terms of order $\mathcal{O}(\sigma)$ cancel and we arrive at $\langle f \rangle' = \langle f \rangle + \mathcal{O}(\sigma^2)$.

### D. $r$-velocity fluctuations

The crucial part of our ansatz is the transformation for the $r$-velocity fluctuations mapping $\mathbf{v}(\mathbf{r}, t|\mathbf{s})$ to $\mathbf{v}'(\mathbf{r}', t'|\mathbf{s}')$. This shall be realized through a linear transformation operator $\mathcal{T}$ which expresses the abstract elimination of degrees of freedom and the subsequent rescaling:

$$\mathbf{v}' = \mathbf{v} + \sigma \mathcal{T}[\mathbf{v}] + \mathcal{O}(\sigma^2). \tag{28}$$

The linearity of $\mathcal{T}$ avoids terms more-than-quadratic in $\mathbf{v}$ in the equation of motion. $\mathcal{T}$ will in general not be a simple rescaling, but may transform $\mathbf{v}$ and its derivatives differently. This takes care of the possibility that different orders of turbulent velocity derivatives might be independent scaling fields. Such additional scaling fields would be analogous to the additional order parameters in dynamical critical phenomena introduced by conserved quantities relevant on the large scales.

Of course, we have to specify an explicit expression for $\mathcal{T}$. It shall cover all linear differential operators of finite order in $\mathbf{r}$, $\mathbf{s}$ and $L$. This choice is not motivated physically.

$$\mathcal{T}_i[\mathbf{v}] = \sum_{n=0}^{N} \sum_{m=0}^{M} \left[ \alpha_{i,j;k_1,\ldots,k_n}^{(n,m)} \partial_{r_{k_1}} \cdots \partial_{r_{k_n}} \partial_L^m + \beta_{i,j;k_1,\ldots,k_n}^{(n,m)} \partial_{s_{k_1}} \cdots \partial_{s_{k_n}} \partial_L^m \right] v_j. \tag{29}$$

The derivatives $\partial_{s_i}$ and $\partial_L$ are defined with $\mathbf{r}$ kept fixed according to

$$\begin{aligned}\boldsymbol{\nabla}_{\mathbf{s}} &\equiv \boldsymbol{\nabla}_{\mathbf{s}}\big|_{\mathbf{r},t,L} = \left(\boldsymbol{\nabla}_{\mathbf{x}_0}\mathbf{s}\big|_{t,L}\right)^{-1} \cdot \boldsymbol{\nabla}_{\mathbf{x}_0}\big|_{\mathbf{r},t,L}, \\ \partial_L &\equiv \partial_L\big|_{\mathbf{r},t,\mathbf{s}} = \partial_L\big|_{\mathbf{r},t,\mathbf{x}_0} - \partial_L \mathbf{s}\big|_{t,\mathbf{x}_0} \cdot \boldsymbol{\nabla}_{\mathbf{s}}\big|_{\mathbf{r},t,L}.\end{aligned} \tag{30}$$

This is well-defined since the uniqueness of the marker trajectories allows to unambiguously rewrite $\mathbf{v}$ as a function of $\mathbf{s}$ instead of $\mathbf{x}_0$. It should be stressed that the time derivative in the equation of motion for $\mathbf{v}$ and $\mathbf{w}$ is not taken with constant $\mathbf{s}$, but $\mathbf{x}_0$, i.e. $\partial_t \equiv \partial_t\big|_{\mathbf{r},\mathbf{x}_0,L}$. The characteristic property of $\mathbf{v}$ (and $\mathbf{w}$) being velocity differences, cf. (8), allows to express $\mathbf{s}$ derivatives through $\mathbf{r}$ derivatives:

$$\partial_{s_i}\mathbf{v} = S[\partial_{r_i}\mathbf{v}], \qquad \partial_{s_i}\partial_{r_j}\mathbf{v} = \partial_{r_i}\partial_{r_j}\mathbf{v}, \qquad \text{etc.} \tag{31}$$

Because of this "difference property" mixed derivatives in $\mathbf{r}$ and $\mathbf{s}$ are not independent and thus omitted from (29). The coefficients $\alpha^{(n,m)}_{i,j;k_1,\ldots,k_n}$ and $\beta^{(n,m)}_{i,j;k_1,\ldots,k_n}$ are (continuously differentiable) functions of $\mathbf{r}$, $L$, and $\mathbf{s}$ (later $s_2$ only). Without loss of generality we can set $\beta^{(0,m)}_{i,j} = 0$ and assume these coefficients to be symmetric in the indices $k_1, \ldots, k_n$. The finiteness of $M$ and $N$ will not be used explicitly, but in order to find or exclude solutions for $\mathcal{T}$ with infinitely many contributions in (29), we would have to study questions of convergence and exchange of limits in a suitably chosen function space.

It seems reasonable that $\mathbf{v}'$ should have the same basic properties as $\mathbf{v}$. Obviously $\mathbf{v}'$ has to be divergence free and has to vanish for $\mathbf{r}' = 0$. We also demand a zero mean of $\mathbf{v}'$ as a consequence of homogeneity. Furthermore, we try to preserve the difference property. But if $\mathbf{u}'$ and $\mathbf{v}'$ decouple as argued in Sect. III B this constraint could or should be relaxed. These four properties lead to corresponding constraints for $\mathcal{T}$:

$$\boldsymbol{\nabla}_\mathbf{r}\cdot\mathcal{T}[\mathbf{v}] = 0 \quad \text{for} \quad \boldsymbol{\nabla}_\mathbf{r}\cdot\mathbf{v} = 0 \qquad \text{(divergence freeness)}, \tag{32}$$

$$\mathcal{T}[\mathbf{v}]\big|_{\mathbf{r}=0} = 0 \quad \text{for} \quad \mathbf{v}(0,t) \equiv 0 \qquad \text{(origin)}, \tag{33}$$

$$\langle\mathcal{T}[\mathbf{v}]\rangle = 0 \quad \text{for} \quad \langle\mathbf{v}\rangle = 0 \qquad \text{(homogeneity)}, \tag{34}$$

$$(S\boldsymbol{\nabla}_\mathbf{r} - \boldsymbol{\nabla}_\mathbf{s})\mathcal{T}[\mathbf{v}] = S\boldsymbol{\nabla}_\mathbf{r}\mathbf{v} \tag{35}$$

$$\text{for} \quad \boldsymbol{\nabla}_\mathbf{s}\mathbf{v} = S[\boldsymbol{\nabla}_\mathbf{r}\mathbf{v}] \qquad \text{(difference characteristic)}.$$

They are not automatically fulfilled by ansatz (29) and will be enforced later (Sect. V A).

### E. Eulerian velocities

Since the Eulerian velocity $\mathbf{u}'$ defined in (26) should follow a Navier–Stokes equation and because of the formal similarity to the equation of motion for $\mathbf{v}$, we employ a transformation $\widetilde{\mathcal{T}}$ for $\mathbf{u}$ analogous to $\mathcal{T}$:

$$\mathbf{u}' = \mathbf{u} + \sigma \widetilde{\mathcal{T}}[\mathbf{u}] + \mathcal{O}(\sigma^2),$$
$$\widetilde{\mathcal{T}}_i[\mathbf{u}] = \sum_{n=0}^{\tilde{N}} \sum_{m=0}^{\tilde{M}} \tilde{\alpha}^{(n,m)}_{i,j;k_1,\ldots,k_n} \partial_{s_{k_1}} \cdots \partial_{s_{k_n}} \partial_L^m u_j. \tag{36}$$

The coefficients $\tilde{\alpha}^{(n,m)}_{i,j;k_1,\ldots,k_n}$ are functions of $\mathbf{s}$ and $L$ and each is symmetric in the indices $k_1,\ldots,k_n$ without loss of generality. Divergence freeness of $\mathbf{u}'$ leads to the constraint

$$\boldsymbol{\nabla}_\mathbf{s} \cdot \widetilde{\mathcal{T}}[\mathbf{u}] = 0 \quad \text{for} \quad \boldsymbol{\nabla}_\mathbf{s} \cdot \mathbf{u} = 0. \tag{37}$$

## V. INVARIANCE OF THE NONLINEARITY

After having specified the transformation, we can apply it to the equation of motion for $\mathbf{v}$. But at first, only the linearity of $\mathcal{T}$ is used, and we introduce the abbreviations $\mathcal{P} = \mathcal{SD}$ and $\mathcal{P}' = \mathcal{SD}'$. After inserting (22), (27) and (28) into the lhs. of (21) and $\partial_t \mathbf{v}$ from (20) we find

$$\partial_{t'} \mathbf{v}' + \mathcal{P}'[\mathbf{v}' \cdot \boldsymbol{\nabla}_{\mathbf{r}'} \mathbf{v}'] = (1 - \sigma z + \sigma \mathcal{T})\mathcal{L}[\mathbf{v}] + \sigma \dot{\mathcal{T}}[\mathbf{v}] + (\mathbf{u}' - (1 - \sigma z)\mathbf{u}) \cdot \boldsymbol{\nabla}_\mathbf{s} \mathbf{v}$$
$$+ \sigma(z - 1 - \mathcal{T})\mathcal{P}[\mathbf{v} \cdot \boldsymbol{\nabla}_\mathbf{r} \mathbf{v}] + \sigma \delta \mathcal{P}[\mathbf{v} \cdot \boldsymbol{\nabla}_\mathbf{r} \mathbf{v}] + \sigma \mathcal{P}\big[\mathcal{T}[\mathbf{v}] \cdot \boldsymbol{\nabla}_\mathbf{r} \mathbf{v} + \mathbf{v} \cdot \boldsymbol{\nabla}_\mathbf{r} \mathcal{T}[\mathbf{v}]\big] + \mathcal{O}(\sigma^2)$$
$$\stackrel{!}{=} \mathcal{L}'[\mathbf{v}'] + \mathcal{O}(\sigma^2).$$

Here, $\delta \mathcal{P}$ is defined through $\mathcal{P}' = \mathcal{P} + \sigma\,\delta\mathcal{P} + \mathcal{O}(\sigma^2)$ and $\dot{\mathcal{T}}$ in an abbreviation for $\partial_t \mathcal{T} - \mathcal{T} \partial_t$. The nonlinear terms have to cancel out, since $\mathcal{L}'$ is supposed to be inhomogeneous linear in $\mathbf{v}$:

$$\mathcal{P}\big[\mathcal{T}[\mathbf{v}] \cdot \boldsymbol{\nabla}_\mathbf{r} \mathbf{v} + \mathbf{v} \cdot \boldsymbol{\nabla}_\mathbf{r} \mathcal{T}[\mathbf{v}]\big] - \mathcal{T}\mathcal{P}[\mathbf{v} \cdot \boldsymbol{\nabla}_\mathbf{r} \mathbf{v}] = -((z-1)\mathcal{P} + \delta\mathcal{P})[\mathbf{v} \cdot \boldsymbol{\nabla}_\mathbf{r} \mathbf{v}]. \tag{38}$$

This inhomogeneous linear constraint for the operator $\mathcal{T}$ is at the heart of our RG approach for turbulence. We call such a constraint the "nonlinearity constraint". From the remaining terms and because of $\mathcal{L}'[\mathbf{v}'] = \mathcal{L}'[\mathbf{v}] + \sigma \mathcal{L}\mathcal{T}[\mathbf{v}] + \mathcal{O}(\sigma^2)$ we infer an equation which determines $\mathcal{L}'$:

$$\mathcal{L}'[\mathbf{v}] = (1 - \sigma z)\mathcal{L}[\mathbf{v}] + \sigma(\mathcal{T}\mathcal{L} - \mathcal{L}\mathcal{T})[\mathbf{v}] + \sigma\dot{\mathcal{T}}[\mathbf{v}] + (\mathbf{u}' - (1 - \sigma z)\mathbf{u})\cdot\boldsymbol{\nabla}_\mathbf{s}\mathbf{v} + \mathcal{O}(\sigma^2). \quad (39)$$

It will be analyzed in Sect. VI. The inhomogeneous linearity of $\mathcal{L}'$ will then be confirmed a posteriori.

We were not able to find the general solution of the nonlinearity constraint (38) within the ansatz (29). The main difficulties arise from the nonlocality of the projector $\mathcal{P}$. We therefore first look for solutions which fulfill a simpler constraint, namely we ignore the pressure term. The resulting modification of the nonlinearity constraint allows us to find its general solution. Formally, this means to substitute the identity operator for $\mathcal{P}$ and to set $\delta\mathcal{P}$ to zero in (38). This leads to the considerably simpler constraint

$$\mathcal{T}[\mathbf{v}]\cdot\boldsymbol{\nabla}_\mathbf{r}\mathbf{v} + \mathbf{v}\cdot\boldsymbol{\nabla}_\mathbf{r}\mathcal{T}[\mathbf{v}] - \mathcal{T}[\mathbf{v}\cdot\boldsymbol{\nabla}_\mathbf{r}\mathbf{v}] = -(z-1)\mathbf{v}\cdot\boldsymbol{\nabla}_\mathbf{r}\mathbf{v}. \quad (40)$$

In a second step, we demand that the solution $\mathcal{T}$ of (40) meets the full nonlinearity constraint (38). Thus, the effects of pressure are only taken care of subsequently. For this second step it is *sufficient* that

$$(\mathcal{T}\mathcal{P} - \mathcal{P}\mathcal{T})[\mathbf{q}] = \delta\mathcal{P}[\mathbf{q}] \qquad \text{for} \qquad \mathbf{q} = \mathbf{v}\cdot\boldsymbol{\nabla}_\mathbf{r}\mathbf{v}. \quad (41)$$

In general, this constraint will reduce the set of solutions. The geometry of the flow volume enters only here. One should keep in mind that this two-stage method might not find all solutions of (38). For example, there might exist additional solutions where the pressure term and the local nonlinear term "mix" or the equality in (40) holds only except for a divergence free, additive term.

## A. Neglecting pressure

Determining the solutions of (40) is quite tedious and boring. The lhs. is calculated for all terms of the sum in ansatz (29). It is assumed that all components of **v** and all their derivatives are independent from each other — except for derivatives which are coupled because of the divergence freeness and must be reexpressed through independent derivatives. Then, (40) must be individually fulfilled by all terms with a common derivative structure. More calculational details of this straightforward procedure can be found in the Appendix A. Here, we proceed to the result: The general solution of (40) may be written in the form

$$\mathcal{T}_i[\mathbf{v}] = \alpha_{i,j}\, v_j - [(z-1)r_j + \alpha_{j,k}\, r_k - \gamma_j](\partial_{r_j} v_i) + \beta_j \partial_{s_j} v_i + \beta_L \partial_L v_i \tag{42}$$

with 17 free parameters $\alpha_{i,j}$, $\gamma_j$, $\beta_j$, $\beta_L$, and $z$, which are arbitrarily depending on **s** and $L$ (except for $z = \text{const.}$). The $\alpha_{i,j}$ are dimensionless, whereas $\gamma_j$, $\beta_j$ and $\beta_L$ carry the dimension of a length. It should be noted that divergence freeness is not relevant for this result: The same general solution holds if one assumes *all* derivatives to be independent of each other.

Now we have to check the constraints (32)–(35). The divergence constraint (32) and the homogeneity constraint (34) are automatically fulfilled by all transformations (42). The origin constraint (33) demands $\gamma_j = 0$ for all $j = 1,2,3$. The difference constraint (35) leads to

$$(S\partial_{r_l} - \partial_{s_l})\mathcal{T}_i[\mathbf{v}] = -[(z-1)\delta_{jl} + \alpha_{j,l} + (\partial_{s_l}\beta_j)]S\partial_{r_j} v_i + \gamma_j \partial_{r_j}\partial_{r_l} v_i\big|_{\mathbf{r}=0}$$
$$- \big\{(\partial_{s_l}\alpha_{i,j})v_j - [\partial_{s_l}((z-1)\delta_{jk} + \alpha_{j,k})]r_k\, \partial_{r_j} v_i + (\partial_{s_l}\gamma_j)\partial_{r_j} v_i + (\partial_{s_l}\beta_L)\partial_L v_i\big\}$$
$$\stackrel{!}{=} S\partial_{r_l} v_i.$$

The terms in curly brackets all have a different structure and must vanish individually. Consequently, all coefficient functions except $\beta_j$ must be independent of **s**, i.e., they are free constant numbers, which is compatible with the constantness of $z$. From the remaining terms we find again $\gamma_j = 0$ and a linear dependence of $\beta_j$ on **s**:

$$\beta_j = -z\, s_j - \alpha_{j,l}\, s_l + \beta_{0;j} \quad \text{with} \quad \beta_{0;j} = \text{const.} \tag{43}$$

We see that the difference constraint restricts the admitted transformations considerably. This goes as far as enforcing a dependence on all components of **s** which contradicts our desire of a sole $s_2$ dependence. Later on, we will therefore also consider weaker forms of the difference constraint. In the extreme case of ignoring this constraint we have to generalize the ansatz (29) to include mixed derivatives with respect to **s** and **r**. But this does not lead to a more general solution (42); the difference constraint does not change the solutions for $\mathcal{T}$ in the pressureless case.

*Supplement: anisotropic rescaling* Imagining a definite RG transformation with elimination of degrees of freedom in a thin boundary layer, an anisotropic rescaling of spatial distances seems much more appropriate than the isotropic rescaling adopted from critical phenomena. We therefore repeat the preceding calculations with

$$r'_i = r_i + \sigma R_{i,j} r_j + \mathcal{O}(\sigma^2), \quad R_{i,j} = \text{const.},$$

instead of (22). Since $\sigma$ is only determined up to an arbitrary constant factor, we may choose to fix one of the parameters $R_{i,j}$, e.g. $R_{2,2} = 1$. This general ansatz includes the physically motivated case $R_{i,j} = \delta_{i,2}\delta_{j,2}$. We indeed find a solution generalizing (A5):

$$\alpha_l^{(1)} = -(\alpha_{l,j}^{(0,0)} - R_{j,l})r_j - z\, r_l + \gamma_l \quad \text{with } \gamma_l \text{ independent of } \mathbf{r}.$$

But the full solution for $\mathcal{T}$ corresponding to (42) complies with the divergence constraint (32) only for $R_{i,j} = \delta_{i,j}$. We conclude that the isotropic nature of incompressibility enforces an isotropic rescaling of **r**, at least for the set of transformations found so far.

*Supplement: spatially varying z exponent* A crucial but perhaps too restrictive assumption is the necessity of only a single dynamical exponent $z$, which might be a speciality of critical phenomena and not valid for turbulence. We therefore might try to use a position dependent exponent $z(\mathbf{s}(t)/L)$ exploiting the fact that there seems to be no quantity analogous to **s** in critical phenomena. Without elaborating on all details (cf. Ref. [60]) here we just mention that such a generalized $z$ is compatible with the transformations $\mathcal{T}$ considered so far, but implies homogeneity and difference constraints which cannot be trivially satisfied.

The main problem, though, comes from the nonlinearity constraint upon the transformation $\widetilde{\mathcal{T}}$ for **u** which will be dealt with in Sect. VI B. It enforces $z = \text{const.}$, and so we shall not pursue this idea further.

### B. Invariance of the pressure term

Now we perform the second step where we have to make sure the commutator constraint (41) holds. One verifies very easily that $\mathcal{T}$ commutes with $\mathcal{S}$ (since $\gamma_j = 0$), and it suffices to check to exchangeability of $\mathcal{T}$ with $\mathcal{D}$ under the action of $\mathcal{S}$:

$$\mathcal{S}(\mathcal{T}\mathcal{D} - \mathcal{D}\mathcal{T} - \delta\mathcal{D})[\mathbf{q}] = 0 \qquad \text{for} \qquad \mathbf{q} = \mathbf{v} \cdot \boldsymbol{\nabla}_\mathbf{r} \mathbf{v}. \tag{44}$$

Here, we define $\delta\mathcal{D}$ through $\mathcal{D}' = \mathcal{D} + \sigma\,\delta\mathcal{D} + \mathcal{O}(\sigma^2)$. After integration by parts we find from (12) and (42) for any Green's function $G_\Omega$

$$\begin{aligned}(\mathcal{T}\mathcal{D} - \mathcal{D}\mathcal{T} - \delta\mathcal{D})_i[\mathbf{q}] = -\int_{\Omega(\mathbf{s})} d^3\boldsymbol{\rho} &\Big\{ (\alpha_{i,j} + \alpha_{j,i})\big(\partial_{r_j} G_\Omega(\mathbf{r}+\mathbf{s}, \boldsymbol{\rho}+\mathbf{s})\big) \\
&- \partial_{r_i}\Big[\Big(z\,(r_j \partial_{r_j} + \rho_j \partial_{\rho_j} + 1) + \alpha_{j,k}\,(r_k \partial_{r_j} + \rho_k \partial_{\rho_j} + \delta_{j\,k}) \\
&\qquad - \beta_j \partial_{s_j} - \beta_L \partial_L\Big) G_\Omega(\mathbf{r}+\mathbf{s}, \boldsymbol{\rho}+\mathbf{s})\Big]\Big\} \big(\partial_{\rho_l} q_l(\boldsymbol{\rho}, t)\big). \end{aligned} \tag{45}$$

This expression can now be evaluated for different Green's functions corresponding to different geometries. Although we would like to focus on plane shear flow because of its physical significance, we will also study an "infinite" system which is similar to what is considered in most other theories of turbulence. This strategy enables us to get insight about the consequences of the presence of boundaries.

Let us first consider the infinite system's Green's Function,

$$G_\Omega(\mathbf{x}, \boldsymbol{\xi}) \mapsto G_\mathrm{f}(\mathbf{x}, \boldsymbol{\xi}) = -\frac{1}{4\pi} \frac{1}{|\mathbf{x} - \boldsymbol{\xi}|}. \tag{46}$$

We prefer to call this substitution an "approximation" to the plane shear case since a truly infinite turbulent system does not make sense within our framework with its physical forcing. Effects due to the boundary conditions and the finiteness in one space direction are neglected. Therefore, we keep the transformation (25) of $\mathbf{s}$ for the infinite system although the reasoning for it is not transferable since $\mathcal{D}$ is independent of $\mathbf{s}$. With the help of the relation

$$(r_j\partial_{r_j} + \rho_j\partial_{\rho_j} + 1)G_\mathrm{f}(\mathbf{r}, \boldsymbol{\rho}) = 0 \tag{47}$$

the commutator (45) can be evaluated as

$$(\mathcal{T}\mathcal{D} - \mathcal{D}\mathcal{T} - \delta\mathcal{D})_i[\mathbf{q}] = -\int_{\Omega(\mathbf{s})} d^3\boldsymbol{\rho} \left\{ 3\sum_{j<k}(\alpha_{j,k} + \alpha_{k,j})\frac{(r_k - \rho_k)(r_j - \rho_j)}{|\mathbf{r} - \boldsymbol{\rho}|^2} \right.$$
$$\left. - \alpha_{j,j}\left(1 - 3\frac{(r_j - \rho_j)^2}{|\mathbf{r} - \boldsymbol{\rho}|^2}\right)\right\} (\partial_{r_i}G_\mathrm{f}(\mathbf{r}, \boldsymbol{\rho}))(\partial_{\rho_l}q_l(\boldsymbol{\rho}, t)).$$

The commutator constraint (44) holds only if

$$\alpha_{j,k} = -\alpha_{k,j} \quad \text{for} \quad j \neq k \quad \text{and} \quad \alpha_{j,j} \quad \text{independent of} \quad j. \tag{48}$$

This expresses the rotational symmetry of the problem.

Translational invariance and the absence of a length scale set by the boundary conditions are responsible for the independence of $G_\mathrm{f}(\mathbf{r} + \mathbf{s}, \boldsymbol{\rho} + \mathbf{s})$ from $\mathbf{s}$ and $L$. We demand this independence also for the transformation $\mathcal{T}$:

$$\alpha_{i,j}, z \quad \text{independent of} \quad \mathbf{s}, L \quad \text{and} \quad \beta_j = \beta_L = 0.$$

This contradicts the difference constraint (43) (except for the uninteresting case $\alpha_{i,j} = z = 0$). In order to avoid $\mathbf{s}$ dependent transformations, one has to give up the difference constraint. This seems reasonable with respect to the decoupling of $\mathbf{v}$ and $\mathbf{u}$ discussed in Sect. III B. Then, the general solution (42) may be written in the form

$$\mathcal{T}_i[\mathbf{v}] = \alpha_{i,j}\,v_j - \alpha_1 r_j\partial_{r_j}v_i - \frac{1}{2}\sum_{j\neq k}\alpha_{j,k}\,(r_k\partial_{r_j} - r_j\partial_{r_k})v_i, \tag{49}$$

$$\alpha_1 = z - 1 + \alpha_0, \qquad \alpha_0 = \alpha_{j,j}, \tag{50}$$

with 5 independent, constant and dimensionless parameters $\alpha_{1,2}$, $\alpha_{1,3}$, $\alpha_{2,3}$, $\alpha_1$, and $z$.

Now we turn to the physically relevant case of plane shear flow. For its Green's function $G_\Omega$, Eq. (10),

$$(r_j\partial_{r_j} + \rho_j\partial_{\rho_j} + s_2\partial_{s_2} + L\partial_L + 1)G_\Omega(\mathbf{r}+\mathbf{s}, \boldsymbol{\rho}+\mathbf{s}) = 0 \tag{51}$$

holds instead of (47). The complete evaluation of the commutator constraint (45) leads to a rather complicated expression for which it is difficult to find all solutions. But we find (45) to be satisfied by all transformations $\mathcal{T}$ with

$$\begin{gathered}\alpha_{1,2} = \alpha_{2,1} = \alpha_{2,3} = \alpha_{3,2} = 0, \quad \alpha_{1,3} = -\alpha_{3,1},\\ \alpha_1 \text{ as in (50)}, \quad \beta_2 = -(1+\alpha_1)s_2, \quad \beta_L = -(1+\alpha_1)L.\end{gathered} \tag{52}$$

For this, one only needs to know that $G_\Omega$ has to adopt the form $G_\Omega(\mathbf{r}+\mathbf{s}, \boldsymbol{\rho}+\mathbf{s}) = \widetilde{G}_\Omega\big((\mathbf{r}+\mathbf{s})/L, (\boldsymbol{\rho}+\mathbf{s})/L\big)/L$ because of dimensional reasons and that $G_\Omega$ depends on $r_1$, $\rho_1$, $r_3$, and $\rho_3$ only in the combination $(r_1-\rho_1)^2 + (r_3-\rho_3)^2$ because of translational invariance in planes parallel to the plates.

Due to this symmetry we additionally demand that the transformation should not depend on $s_1$ and $s_3$ as discussed before. This allows for explicitly space dependent transformations with

$$\alpha_{i,j}(s_2/L) \text{ independent of } s_1, s_3 \text{ and } \beta_1 = \beta_3 = 0.$$

But this space dependence again contradicts the difference constraint. Once more, we decide to give up this constraint in favour of obtaining a different set of admissed transformations which can be written in the form

$$\begin{aligned}\mathcal{T}_i[\mathbf{v}] = {}& \alpha_0\,v_i + \alpha_{1,3}(\delta_{i,1}v_3 - \delta_{i,3}v_1) \\ & - \alpha_1 r_j\partial_{r_j}v_i - \alpha_{1,3}(r_3\partial_{r_1} - r_1\partial_{r_3})v_i - (1+\alpha_1)(s_2\partial_{s_2} + L\partial_L)v_i\end{aligned} \tag{53}$$

with 3 independent and dimensionless parameters, namely $\alpha_{1,3}$ and $\alpha_1$, depending on $s_2/L$, and constant $z$. Unfortunately, we can not come up with a satisfactory physical interpretation of this transformation yet.

# VI. TRANSFORMATION OF DISSIPATION AND FORCING

Having found sets of admitted transformations, (49) and (53) respectively, we transform the linear part $\mathcal{L}'$ of the equation of motion for $\mathbf{v}$.

## A. Viscosity

We begin with the calculation of $(1 - \sigma z)\mathcal{L} + \sigma(\mathcal{T}\mathcal{L} - \mathcal{L}\mathcal{T})$ in (39) for the viscous part $\nu\mathcal{S}\triangle_{\mathbf{r}}$ of $\mathcal{L}$ in the case of plane shear flow:

$$(1 - \sigma z)\nu\mathcal{S}\triangle_{\mathbf{r}}\mathbf{v} + \sigma(\mathcal{T}\nu\mathcal{S}\triangle_{\mathbf{r}} - \nu\mathcal{S}\triangle_{\mathbf{r}}\mathcal{T})\mathbf{v} = (1 + \sigma(2\alpha_1 - z))\nu\mathcal{S}\triangle_{\mathbf{r}}\mathbf{v} + \mathcal{O}(\sigma^2).$$

Taking into account the rescaling of $\triangle_{\mathbf{r}}$ we find $\nu'\triangle_{\mathbf{r}'}\mathbf{v}'$ to be a renormalized viscous contribution to $\mathcal{L}'[\mathbf{v}']$ with the rescaled viscosity

$$\nu' = (1 + \zeta_\nu \sigma)\nu + \mathcal{O}(\sigma^2) \quad \text{with} \quad \zeta_\nu = 2(1 + \alpha_1) - z. \tag{54}$$

Here, the symmetry properties of the parameters $\alpha_{i,j}$, cf. (48) and (52), have to be used explicitly. This result also holds for $z(s_2/L)$ and also for the infinite system approximation. In the former case $\nu$ has to be generalized to a function of $s_2/L$.

Since the viscous part of $\mathcal{L}$ is reproduced with exactly the same structure, we can safely regard all other terms in (39) as contributions to the renormalized forcing $\mathcal{F}'$. The property $\nu' > \nu$ of an inverse RG transformation implies the constraint $\zeta_\nu > 0$.

## B. Forcing

The admitted transformations $\widetilde{\mathcal{T}}$ for the Eulerian velocity $\mathbf{u}$ which are covered by the ansatz (36) can be deduced easily from the results of Sect. V because of the formal similarity of the Navier–Stokes equation and the equation of motion for $\mathbf{v}$:

$$\widetilde{\mathcal{T}}_i[\mathbf{u}] = \tilde{\alpha}_{i,j}\, u_j - [z\, s_j + \tilde{\alpha}_{j,k}\, s_k](\partial_{s_j} u_i) + \tilde{\beta}_L \partial_L u_i. \tag{55}$$

The parameters $\tilde{\alpha}_{j,k}$ and $\tilde{\beta}_L$ are geometry dependent in the same way as $\alpha_{i,j}$ and $\beta_L$, but in any case have to be independent of $\mathbf{s}$. In general, the viscosity in the Navier–Stokes equation rescales with a different scaling factor and therefore will be denoted with the new symbol $\tilde{\nu}$:

$$\tilde{\nu}' = (1 + \tilde{\zeta}_\nu \sigma)\tilde{\nu} + \mathcal{O}(\sigma^2) \quad \text{with} \quad \tilde{\zeta}_\nu = 2(1 + \tilde{\alpha}_1) - z. \tag{56}$$

Next, we calculate the last two terms in (39). First we look at $\dot{\mathcal{T}}$ for plane shear flow. In the most simple case of constant parameters $\alpha_0$, $\alpha_1$, and $\alpha_{1,3}$ we find from (53)

$$\dot{\mathcal{T}} = \partial_t \mathcal{T} - \mathcal{T}\partial_t = -(1 + \alpha_1)\left[u_2 \partial_{s_2} - s_2(\partial_{s_2}\mathbf{u})\cdot\boldsymbol{\nabla}_\mathbf{s} - L(\partial_L \mathbf{u})\cdot\boldsymbol{\nabla}_\mathbf{s}\right]. \tag{57}$$

Here, we made use of the commutators $\partial_t \boldsymbol{\nabla}_\mathbf{s} - \boldsymbol{\nabla}_\mathbf{s}\partial_t = -(\boldsymbol{\nabla}_\mathbf{s}\mathbf{u})\cdot\boldsymbol{\nabla}_\mathbf{s}$ and $\partial_t \partial_L - \partial_L \partial_t = -(\partial_L \mathbf{u})\cdot\boldsymbol{\nabla}_\mathbf{s}$ which follow from definition (30). We see that $\sigma\dot{\mathcal{T}}$ has the same structure as $(\mathbf{u}' - (1 - \sigma z)\mathbf{u})\cdot\boldsymbol{\nabla}_\mathbf{s}$ in (39), but these two contributions to $\mathcal{F}'$ have a different structure than the remaining contribution $\sigma(\mathcal{T}\mathcal{F} - \mathcal{F}\mathcal{T})$, since the former do not contain the turbulent profile $\mathbf{U}$. Thus, they describe structurally new, additional forces (per mass). Since such an additional forcing is compatible with the idea of our RG approach, one should study how it behaves under the RG transformations' next iteration. We now show for one specific example that structurally new terms appear with each RG iteration. A term in (57) like $s_2(\partial_{s_2}\mathbf{u})\cdot\boldsymbol{\nabla}_\mathbf{s}$ in $\mathcal{F}$ results in a contribution

$$-\sigma(1 + \alpha_1)(s_2)^2\left[(\partial_{s_2}\partial_{s_2}\mathbf{u})\cdot\boldsymbol{\nabla}_\mathbf{s} - (\partial_{s_2}\mathbf{u})\cdot\boldsymbol{\nabla}_\mathbf{s}\,\partial_{s_2}\right]$$

to $\mathcal{F}'$ from $\sigma(\mathcal{T}\mathcal{F} - \mathcal{F}\mathcal{T})$. One easily sees that with each iteration a new contribution with an additional $s_2$ derivative is created. The same holds true for nonconstant $\alpha$ parameters which make appear even more distinct forcing terms. One might try to somehow get rid of this additional forcing. We thoroughly analyzed several possibilities which all resulted in the necessity to give up one or the other fundamental requirement [60]. We conclude that the appearance of additional forcing terms is inevitable, although we could not further analyze the transformation behaviour.

We are now able to calculate the turbulent profile, that is the mean value of the transformation (55) for **u**. For plane shear flow we have $U_2' = 0$ and

$$U_i' = U_i + \sigma\{\tilde{\alpha}_{i,j} U_j - (\tilde{\alpha}_0 + z)x_2 \partial_{x_2} U_i + \tilde{\beta}_L \partial_L U_i\} + \mathcal{O}(\sigma^2), \quad (i = 1, 3). \tag{58}$$

Unfortunately, this equation does not allow a unique determination of the profile. But all profiles which take on the functional form $U^* = U_L f(x_2/L)$ and are bounded for all $x_2$ and $\nu \to 0$ are admissible solutions of (58) (with $\tilde{\alpha}_0 = 0$). In the remaining part of this work, we will consider only this case $\tilde{\alpha}_0 = 0$.

Finally we are in the position to discuss the total transformation of the forcing $\mathcal{F}$. The calculation of the remaining contribution $(1 - \sigma z)\mathcal{F} + \sigma(\mathcal{T}\mathcal{F} - \mathcal{F}\mathcal{T})$ in (39) shows that $\mathcal{F}$ would be structurally reproduced (except for the additional forcing discussed above), if

$$U_i' = U_i + \sigma\{\alpha_{i,j} U_j - (\alpha_0 + z)x_2 \partial_{x_2} U_i + \beta_L \partial_L U_i\} + \mathcal{O}(\sigma^2) \quad (i = 1, 3)$$

would hold. Since this is not true (cf. (58)) yet another, **U** dependent type of additional forces appears.

We conclude that a closed, iterable RG equation for the forcing $\mathcal{F}$ does not seem to exist. Clearly, more work on the transformation properties of the forcing is necessary. For example, one would have to examine if the additional forces are irrelevant in the sense of RG theory. Also, a restriction of **U** such that the iteration becomes closed could be possible.

## VII. ENERGY DISSIPATION AND STRUCTURE FUNCTIONS

We are done with the transformation of the equations of motion for **v** and **u**. This enables us to look how physical quantities, more specifically energy dissipation and structure functions, behave under the set of admitted RG transformations for **v**, namely (49) with the parameters $\alpha_{1,2}$, $\alpha_{1,3}$, $\alpha_{2,3}$, $\alpha_1$, and $z$ for the infinite system and (53) with the parameters $\alpha_{1,3}$, $\alpha_1$, and $z$ for plane shear flow, respectively.

**A. Invariance of energy dissipation**

The transformed turbulent dissipation rate $\varepsilon_t$ is obtained from the representation (18) and the RG transformation (42), depending on the choice for the parameters $\alpha_{i,j}$, $\beta_j$, and $\beta_L$.

For the infinite system approximation our requirement of structural invariance of $\varepsilon_t$ is fulfilled automatically, except for a simple rescaling:

$$\varepsilon_t'(\nu') = (1 + \zeta_\varepsilon \sigma)\, \varepsilon_t(\nu) + \mathcal{O}(\sigma^2) \quad \text{with} \quad \zeta_\varepsilon = \zeta_\nu + 2(\alpha_0 - 1 - \alpha_1). \tag{59}$$

Structural invariance is most naturally satisfied by setting $\zeta_\varepsilon = 0$, and this also keeps the value of $\varepsilon_t$ the same:

$$\varepsilon_t' = \varepsilon_t'(\nu') = \varepsilon_t(\nu') = \varepsilon_t(\nu) = \varepsilon_t.$$

Because of our second invariance constraint for the energy dissipation, $\varepsilon' = \varepsilon$, the value of the profile's dissipation is invariant as well, $\varepsilon_U' = \varepsilon_U$. But we do not have any knowledge about the functional structure of $\varepsilon_U'(\nu')$ as long as we do not know how to handle the transformation of the forcing.

For plane shear flow the structure of $\varepsilon_t$ does not reproduce itself:

$$\varepsilon_t'(\nu', x_2, L, \mathbf{U}') = [1 + \zeta_\varepsilon \sigma - \sigma(1 + \alpha_1)(x_2 \partial_{x_2} + L\partial_L)]\,\varepsilon_t(\nu, x_2, L, \mathbf{U}) + \mathcal{O}(\sigma^2). \tag{60}$$

The contributions in addition to (59) are interpreted as nonuniversal since they are related to the boundaries of the system, and are therefore put into $\varepsilon_U'$:

$$\varepsilon_U' = \varepsilon_U'(\nu', x_2, L, \mathbf{U}') + \sigma(1 + \alpha_1)(x_2 \partial_{x_2} + L\partial_L)\,\varepsilon_t(\nu, x_2, L) + \mathcal{O}(\sigma^2).$$

Again we set $\zeta_\varepsilon = 0$ to guarantee structural invariance of $\varepsilon_t$, i.e., $\varepsilon_t(\nu') = \varepsilon_t(\nu)$.

In both cases the condition $\zeta_\varepsilon = 0$ serves as an additional constraint for the RG transformations $\mathcal{T}$ which allows to eliminate one parameter which we choose to be $\alpha_0$. Using $\zeta_\nu$ from (54) and $\alpha_0$ from (50) we find $\alpha_0 = \frac{1}{3}(1 + \alpha_1)$ and $z = \frac{2}{3}(1 + \alpha_1)$ with the only remaining free parameter $\alpha_1$. This value for $z$ gives us $\zeta_\nu = \frac{4}{3}(1 + \alpha_1)$, which is different from $\tilde{\zeta}_\nu = \frac{2}{3}(1+\alpha_1)$. This means that $\nu$, the kinematic viscosity damping the $r$-velocity fluctuations, and $\tilde{\nu}$, responsible for viscous losses of the Euler field, scale differently.

### B. Scaling behaviour of structure functions

Finally we come to the transformation of the structure functions $D^{(m)}$, defined in (14), and their inertial range scaling. Contrary to $\varepsilon_t$ we demand the same functional form for plane shear flow and the infinite system, i.e., $D^{(m)\prime}(\mathbf{r}', \nu', x_2, L) = D^{(m)}(\mathbf{r}', \nu', x_2, L)$, thus not allowing for an additive renormalization.

Again we start with the infinite system and find for the set of admitted transformations

$$D^{(m)}(\mathbf{r}', \nu') = \{1 + \sigma[m\,\alpha_0 - (\alpha_1 r_j + \sum_{k \neq j} \alpha_{j,k}\, r_k)\partial_{r_j}]\}D^{(m)}(\mathbf{r}, \nu) + \mathcal{O}(\sigma^2). \tag{61}$$

We seek for scaling solutions to this RG equation in the limit $\nu \to 0$ which obey the geometric symmetries of the infinite system, i.e., depend on $r = |\mathbf{r}|$ only. Then, (61) simplifies to

$$D^{(m)}(r', \nu') = \{1 + \sigma[m\,\alpha_0 - \alpha_1 r \partial_r]\}D^{(m)}(r, \nu) + \mathcal{O}(\sigma^2).$$

Because $L$ is assumed to be the only scale for intermittency corrections, only solutions are sought which are bounded for $\nu \to 0$ with $\mathbf{r} \neq 0$ fixed. Using $r' = (1+\sigma)r$ on the lhs., the unique solution then is

$$D^{(m)}(r) \propto r^{\zeta_m} \quad \text{with} \quad \zeta_m = \frac{m\alpha_0}{1+\alpha_1} = \frac{m}{3}. \tag{62}$$

It is not surprising that K41 behaviour is enforced since the admitted RG transformations for $\mathbf{v}$ turned out to be independent of $L$. But this result at least confirms consistency of all assumptions behind our method.

For plane shear flow the RG equation

$$D^{(m)}(\mathbf{r}', \nu', x_2, L) = \{1 + \sigma[m\,\alpha_0 - (\alpha_1 r_j + \sum_{j \neq 2, k \neq 2} \alpha_{j,k}\, r_k)\partial_{r_j}$$
$$- (1+\alpha_1)(x_2 \partial_{x_2} + L\partial_L)]\}D^{(m)}(\mathbf{r}, \nu, x_2, L) + \mathcal{O}(\sigma^2) \tag{63}$$

holds. Here, respecting again the geometric symmetries, we are looking for solutions depending on $r$, $x_2$, and $L$ alone, thus simplifying (63) to

$$D^{(m)}(r', \nu', x_2, L) = \{1 + \sigma[m\,\alpha_0 - \alpha_1 r \partial_r$$
$$- (1 + \alpha_1)(x_2 \partial_{x_2} + L \partial_L)]\} D^{(m)}(r, \nu, x_2, L) + \mathcal{O}(\sigma^2).$$

All solutions bounded for $\nu \to 0$ can be represented in this limit as

$$D^{(m)}(r, x_2, L) \propto r^{m/3} f_\mathrm{C}(r/L, x_2/L), \quad \text{with arbitrary function } f_\mathrm{C}. \tag{64}$$

This allows for scaling solutions of the type $D^{(m)} \propto r^{\zeta_m}$ with arbitrary inertial range exponents $\zeta_m$. Within our method we could not yet identify more constraints which would restrict the exponents or even nail them down to a specific value.

## VIII. SUMMARY AND DISCUSSION

### A. Summary of the calculations

The results of Sects. IV to VII can be summarized as follows: with a quite general ansatz for the transformation $\mathcal{T}$ for the $r$-velocity fluctuation $\mathbf{v}$ we evaluate the nonlinearity constraint originating in the invariance of the nonlinear term of the equation of motion. First this is done ignoring the pressure term which restricts $\mathcal{T}$ to a class of solutions with 17 parameters which are allowed to depend on $\mathbf{s}(t)$ and $L$ (except for the dynamical exponent $z$ which must be constant). All further conclusions were drawn for this set of transformations. The difference constraint which expresses that $\mathbf{v}$ is a difference of two Eulerian velocities is abandoned since it turns out to be incompatible with a dependence of $\mathcal{T}$ on $\mathbf{s}(t)$ according to geometric symmetries. This is somewhat surprising and means that the renormalized $r$-velocity fluctuations decouple from the Euler field to become quantities of their own significance. Taking into account a commutator constraint for $\mathcal{T}$ which guarantees the structural invariance of the pressure term and some other constraints restricts the set of solutions further. The number of free parameters is reduced to five constants $\alpha_{1,2}$, $\alpha_{1,3}$, $\alpha_{2,3}$, $\alpha_1$, and $z$ for the infinite system and to one constant $z$ and two functions $\alpha_{1,3}$ and $\alpha_1$ of $s_2/L$ for plane shear flow.

The invariance of the geometry under renormalization suggests a simple transformation of the marker trajectories which may be loosely referred to as $\mathbf{s}' = \mathbf{s}$. We find a transformation $\widetilde{\mathcal{T}}$ for the Eulerian velocity $\mathbf{u}$ which preserves the structure of the nonlinearity of the Navier–Stokes equation. It introduces four new constant parameters for the infinite system and two new constant parameters for plane shear flow. The RG equation for the turbulent shear profile advises a special choice for $\widetilde{\mathcal{T}}$. The viscous term of both equations of motion for $\mathbf{v}$ and $\mathbf{u}$ reproduces structurally, but with differently rescaled viscosities $\nu$ respectively $\tilde{\nu}$. But for the forcing $\mathcal{F}$ several types of structurally new terms appear whose physical meaning is not yet understood. We accept such additional forcing, but do not examine it further at present. Neither the specification of a closed RG equation for $\mathcal{F}$ nor even its existence can be established as matters stand, and we only consider solutions $\mathcal{T}$ with constant parameters. The second important restriction coming from the invariance of the energy dissipation rate is fulfilled by setting the rescaling exponent of the structurally reproducing part of the turbulent dissipation rate to zero. This eliminates one parameter of $\mathcal{T}$.

The RG equations for the structure functions are examined for inertial range solutions with prescribed symmetries corresponding to the geometry of the flow. For the infinite system only K41 scaling exponents turns out to be allowed, whereas for plane shear flow there are scaling solutions with yet arbitrary exponents. This is an indication that intermittency corrections really require the existence of an outer length scale. But neither can the scaling exponents be calculated with this RG method, at least not in its present stage, nor can a physical mechanism for inertial range intermittency be identified. We also investigated periodic boundary conditions and half-space geometry [60] and found that again non-K41 exponents are allowed, implying that both either a zero $\mathbf{v}$-boundary or a finite distance between the plates are sufficient for the possibility of intermittency corrections.

**B. Discussion of the method**

A main advantage of our RG method is its use of the Navier–Stokes equation and the question arises which of its aspects are essential for our results.

The different physical meaning of Eulerian velocities and $r$-velocity fluctuations can be seen in the difference of the transformations $\widetilde{\mathcal{T}}$ and $\mathcal{T}$. The appearance of **s** in Green's function motivates the choice of the transformation of **s** and helps determining $\beta_2$. But the occurrence of the operator $\mathcal{S}$, which introduces the $r$-increment, cf. (5), in the equation of motion for **v** effectively does not enter the evaluation of the commutator constraint. Altogether one sees that it is not essential to introduce Monin's transformation for the results obtained so far.

The quadratic and first-order gradient nature of the nonlinear term influences directly the admissibility of transformations through the nonlinearity constraint. The tensor structure and the dimensionality have no qualitative input on the RG transformation. Formally, all results also apply for general $d$ dimensional turbulence, but since some assumptions such as the invariance of energy dissipation must not be valid in $d = 1$ and $d = 2$ dimensions the physical significance of the results is questionable. We stick to $d = 3$ in this work. The simplified approach of Sect. V A to solve the nonlinearity constraint and the local ansatz for $\mathcal{T}$ do not take into account the strong nonlocality of pressure. It is one of the consequences of Galilean invariance that the fixed prefactor of the nonlinearity is responsible for the elimination of one of the transformation's parameters, namely $\alpha_1$. Otherwise the scaling exponents would be arbitrary also for the infinite system. Concerning Green's function for plane shear flow, only its symmetry properties were utilized. We consider this as a signature of universality: the results are independent of the detailed nature of the boundary conditions.

Incompressibility does not change the set of solutions found for $\mathcal{T}$ (Sect. V A), but it is exploited in the definition of statistical averaging and enforces an isotropic rescaling of **r** (Sect. V A). The definition of statistical averaging is purely formal, but the microscopic nature of the space average enters. The transformations found fulfill $\langle \mathbf{v} \rangle = 0$, but do not take any special advantage of this.

Now we mention some of the open questions which could be investigated further within this method. Most importantly, the transformation of the forcing should be worked out fully by finding a suitable treatment for the structurally new contributions.

One could also think of looking for strongly nonlocal transformations not included in the ansatz (29). We believe this to be quite difficult within the present framework since the simplified two-stage process for solving the nonlinearity constraint (38) is not reasonable then. But if this is accomplished, it would again allow to consider an anisotropic spatial rescaling, new transformations for **s**, and an **s**-dependent dynamical exponent $z$, to give just a few examples. We cannot rule out that new transformations exhibit non-K41 scaling even for the infinite system, although this does not seem very likely.

Of course, the transformation properties of other physical quantities could be studied. Top candidates are the dissipation length $\ell$ and the energy dissipation correlation function. In principle, the examination of the scaling behaviour for $\nu > 0$ could provide insight to the crossover from the inertial to the dissipative range. Finding $\nu$ dependent transformations should also confirm our assumption of $\nu$ independence in the turbulent limit.

The application of our method to other dynamical systems whose scaling behaviour is better known, such as Burgers' equation, Kraichnan's turbulence model for passive scalars, and the GOY cascade model could be possible, but is not self-evident, since the physical idea and the key invariance assumptions do not seem to be transferable in a meaningful way.

To summarize: although we could not calculate concrete numbers for intermittency exponents, we found apparently non-trivial RG transformations which relate intermittency corrections to the *existence* of an outer length scale. Several open questions remain which should be investigated in the future.

**Acknowledgments**: We thank Detlef Lohse for stimulating discussions and Stefan Thomae for bringing Monin's transformation to our attention.

## APPENDIX A: SOLVING THE NONLINEARITY CONSTRAINT

Here we present more details how to determine the RG transformations found in Sect. V A.

The ansatz (29) is plugged into the lhs. of the weakened nonlinearity constraint (40) and derivations are performed such that all derivatives operate directly on **v** components only. This yields the following contributions to the lhs. of (40):

*coefficient functions* $\alpha_{i,j}^{(0,0)}$:

$$\alpha_{l,j}^{(0,0)} v_j (\partial_{r_l} v_i) + (\partial_{r_l} \alpha_{i,j}^{(0,0)}) v_l v_j \tag{A1}$$

*coefficient functions* $\alpha_{i,j}^{(1,0)}$:

$$\alpha_{l,j;k_1}^{(1,0)} (\partial_{r_{k_1}} v_j)(\partial_{r_l} v_i) + (\partial_{r_l} \alpha_{i,j;k_1}^{(1,0)}) v_l (\partial_{r_{k_1}} v_j) - \alpha_{i,j;k_1}^{(1,0)} (\partial_{r_{k_1}} v_l)(\partial_{r_l} v_j) \tag{A2}$$

*coefficient functions* $\alpha_{i,j}^{(2,0)}$:

$$\alpha_{l,j;k_1,k_2}^{(2,0)} (\partial_{r_{k_1}} \partial_{r_{k_2}} v_j)(\partial_{r_l} v_i) + (\partial_{r_l} \alpha_{i,j;k_1,k_2}^{(2,0)}) v_l (\partial_{r_{k_1}} \partial_{r_{k_2}} v_j)$$
$$- \alpha_{i,j;k_1,k_2}^{(2,0)} [(\partial_{r_{k_1}} \partial_{r_{k_2}} v_l)(\partial_{r_l} v_j) + 2(\partial_{r_{k_1}} v_l)(\partial_{r_{k_2}} \partial_{r_l} v_j)] \tag{A3}$$

*coefficient functions* $\alpha_{i,j}^{(3,0)}$:

$$\alpha_{l,j;k_1,k_2,k_3}^{(3,0)} (\partial_{r_{k_1}} \partial_{r_{k_2}} \partial_{r_{k_3}} v_j)(\partial_{r_l} v_i) + (\partial_{r_l} \alpha_{i,j;k_1,k_2,k_3}^{(3,0)}) v_l (\partial_{r_{k_1}} \partial_{r_{k_2}} \partial_{r_{k_3}} v_j)$$
$$- \alpha_{i,j;k_1,k_2,k_3}^{(3,0)} [(\partial_{r_{k_1}} \partial_{r_{k_2}} \partial_{r_{k_3}} v_l)(\partial_{r_l} v_j) + 3(\partial_{r_{k_1}} \partial_{r_{k_2}} v_l)(\partial_{r_{k_3}} \partial_{r_l} v_j)$$
$$+ 3(\partial_{r_{k_1}} v_l)(\partial_{r_{k_2}} \partial_{r_{k_3}} \partial_{r_l} v_j)] \tag{A4}$$

At this point we see that for given $n \geq 1, m = 0$ there are $n-1$ derivatives in the second term. The same is true for all terms except for the second in the contribution for $n-1$. But in general, i.e., except for $n = 1$, these derivatives are distributed differently among the two **v** components in each term. Thus, $\alpha_{i,j;k_1,\ldots,k_n}^{(n,0)} = $ const. follows for $n \geq 2$. Next, we will show that even $\alpha_{i,j;k_1,\ldots,k_n}^{(n,0)} = 0$ for $n \geq 2$. This is immediately clear for odd $n \geq 3$ since respectively one particular term in the second line of (A4) has a unique derivational structure. For even $n \geq 4$ there are respectively two terms, but with distinct derivational structure, leading again to $\alpha^{(n,0)}_{i,j;k_1,\ldots,k_n} = 0$. For $n = 2$ this only follows after evaluating the more complicated constraint

$$\alpha^{(2,0)}_{l,j;k_1,k_2}(\partial_{r_{k_1}}\partial_{r_{k_2}}v_j)(\partial_{r_l}v_i) = \alpha^{(2,0)}_{i,j;k_1,k_2}[(\partial_{r_{k_1}}\partial_{r_{k_2}}v_l)(\partial_{r_l}v_j) + 2(\partial_{r_{k_1}}v_l)(\partial_{r_{k_2}}\partial_{r_l}v_j)].$$

The situation is different for $n = 1$, where the first and second term of (A2) lead to the constraint

$$\alpha^{(1,0)}_{l,j;k_1}(\partial_{r_{k_1}}v_j)(\partial_{r_l}v_i) = \alpha^{(1,0)}_{i,j;k_1}(\partial_{r_{k_1}}v_l)(\partial_{r_l}v_j).$$

A lengthy evaluation yields the representation $\alpha^{(1,0)}_{i,j;k_1} = \delta_{ij}\alpha^{(1)}_{k_1}$ with some new functions $\alpha^{(1)}_{k_1}$.

Now we turn to (A1), the contribution for $n = 0$. The second term has to vanish, because it is the only one without any derivatives of $\mathbf{v}$, which is only possible for $\alpha^{(0,0)}_{i,j} = \mathrm{const}$. The first term is of the same structure as the remaining second term of (A2) and the rhs. of (40), giving finally the constraint

$$\alpha^{(0,0)}_{l,j}v_j(\partial_{r_l}v_i) + (\partial_{r_l}\alpha^{(1,0)}_{i,j;k_1})v_l(\partial_{r_{k_1}}v_j) = -(z-1)v_l(\partial_{r_l}v_i),$$

whose general solution

$$\alpha^{(1)}_l = -\alpha^{(0,0)}_{l,j}r_j - (z-1)r_l + \gamma_l \quad \text{with } \gamma_l \text{ independent of } \mathbf{r} \tag{A5}$$

we obtain after a few more calculational steps.

*coefficient functions $\alpha^{(0,1)}_{i,j}$:*

$$\alpha^{(0,1)}_{l,j}(\partial_L v_j)(\partial_{r_l}v_i) + (\partial_{r_l}\alpha^{(0,1)}_{i,j})v_l(\partial_L v_j) - \alpha^{(0,1)}_{i,j}(\partial_L v_l)(\partial_{r_l}v_j) \tag{A6}$$

*coefficient functions $\alpha^{(0,2)}_{i,j}$:*

$$\alpha^{(0,2)}_{l,j}(\partial_L^2 v_j)(\partial_{r_l}v_i) + (\partial_{r_l}\alpha^{(0,2)}_{i,j})v_l(\partial_L^2 v_j) - \alpha^{(0,2)}_{i,j}[(\partial_L^2 v_l)(\partial_{r_l}v_j) + 2(\partial_L v_l)(\partial_L \partial_{r_l}v_j)] \tag{A7}$$

Similar to the contributions for $n \geq 2, m = 0$ we find $\alpha^{(0,m)}_{i,j} = 0$ for $m \geq 2$. For $m = 1$, on the other hand, $\alpha^{(0,1)}_{i,j} = \mathrm{const.}$ holds because of the second term in (A6). And because of the other two terms we have even

$$\alpha^{(0,1)}_{i,j} = \delta_{ij}\beta_L \quad \text{with } \beta_L \text{ independent of } \mathbf{r}. \tag{A8}$$

*coefficient functions* $\alpha_{i,j}^{(1,1)}$:

$$\alpha_{l,j;k_1}^{(1,1)}(\partial_{r_{k_1}}\partial_L v_j)(\partial_{r_l} v_i) + (\partial_{r_l}\alpha_{i,j;k_1}^{(1,1)})v_l(\partial_{r_{k_1}}\partial_L v_j)$$
$$- \alpha_{i,j;k_1}^{(1,1)}[(\partial_{r_{k_1}}\partial_L v_l)(\partial_{r_l} v_j) + (\partial_{r_{k_1}} v_l)(\partial_L\partial_{r_l} v_j) + (\partial_L v_l)(\partial_{r_{k_1}}\partial_{r_l} v_j)] \quad (A9)$$

This implies $\alpha_{i,j;k_1,\ldots,k_n}^{(n,m)} = 0$ for $n, m \geq 1$.

*coefficient functions* $\beta_{i,j}^{(1,0)}$:

$$\beta_{l,j;k_1}^{(1,0)}(\partial_{s_{k_1}} v_j)(\partial_{r_l} v_i) + (\partial_{r_l}\beta_{i,j;k_1}^{(1,0)})v_l(\partial_{s_{k_1}} v_j) - \beta_{i,j;k_1}^{(1,0)}(\partial_{s_{k_1}} v_l)(\partial_{r_l} v_j) \quad (A10)$$

*coefficient functions* $\beta_{i,j}^{(2,0)}$:

$$\beta_{l,j;k_1,k_2}^{(2,0)}(\partial_{s_{k_1}}\partial_{s_{k_2}} v_j)(\partial_{r_l} v_i) + (\partial_{r_l}\beta_{i,j;k_1,k_2}^{(2,0)})v_l(\partial_{s_{k_1}}\partial_{s_{k_2}} v_j)$$
$$- \beta_{i,j;k_1,k_2}^{(2,0)}[(\partial_{s_{k_1}}\partial_{s_{k_2}} v_l)(\partial_{r_l} v_j) + 2(\partial_{s_{k_1}} v_l)(\partial_{s_{k_2}}\partial_{r_l} v_j)] \quad (A11)$$

As before we infer $\beta_{i,j;k_1,\ldots,k_n}^{(n,0)} = 0$ for $n \geq 2$. But for $n = 1$ $\beta_{i,j;k_1}^{(1,0)} = $ const. holds because of the second term in (A11), and furthermore we have

$$\beta_{i,j;k_1}^{(0,1)} = \delta_{ij}\beta_{k_1} \quad \text{with } \beta_{k_1} \text{ independent of } \mathbf{r}. \quad (A12)$$

*coefficient functions* $\beta_{i,j}^{(1,1)}$:

$$\beta_{l,j;k_1}^{(1,1)}(\partial_{s_{k_1}}\partial_L v_j)(\partial_{r_l} v_i) + (\partial_{r_l}\beta_{i,j;k_1}^{(1,1)})v_l(\partial_{s_{k_1}}\partial_L v_j)$$
$$- \beta_{i,j;k_1}^{(1,1)}[(\partial_{s_{k_1}}\partial_L v_l)(\partial_{r_l} v_j) + (\partial_{s_{k_1}} v_l)(\partial_L\partial_{r_l} v_j) + (\partial_L v_l)(\partial_{s_{k_1}}\partial_{r_l} v_j)] \quad (A13)$$

It follows $\beta_{i,j;k_1,\ldots,k_n}^{(n,m)} = 0$ for $n, m \geq 1$.

Combining the partial solutions (A5)–(A12) we are now able to write down the general solution (42) of (40).

FIGURES

FIG. 1. Plane shear flow between two plates with distance $L$ moving with relative velocity $U_L$

FIG. 2. Illustration of definition (3) of the velocity increment $\mathbf{w}(\mathbf{r}, t|\mathbf{x}_0, t_0)$